# PLANETARY TERRESTRIAL ANALOGUES LIBRARY PROJECT: 3. CHARACTERIZATION OF SAMPLES WITH MICROMEGA

Damien Loizeau, Cédric Pilorget, François Poulet, Cateline Lantz, Jean-Pierre Bibring, Vincent Hamm, Clément Royer, Henning Dypvik, Agata M. Krzesińska, Fernando Rull, Stephanie C. Werner

Highlights:

Building a multi-techniques spectral library to support planetary surface exploration

Use of the NIR hyperspectral microscope MicrOmega for sample characterization

Hyperspectral images allow surface mineralogical mapping at microscopic scale

MicrOmega analysis compared with NIR point spectrometer and Raman analysis

## Abstract

The PTAL (Planetary Terrestrial Analogues Library) project aims at building and exploiting a database involving several analytical techniques, to help characterizing the mineralogical evolution of terrestrial bodies, starting with Mars. Around 100 natural Earth rock samples have been collected from selected locations to gather a variety of analogues for Martian geology, from volcanic to sedimentary origin with different levels of alteration.

All samples are to be characterized within the PTAL project with different mineralogical and elemental analysis techniques, including techniques brought on actual and future instruments at the surface of Mars (Near InfraRed spectroscopy, Raman spectroscopy and Laser Induced Breakdown Spectroscopy).

This paper presents the NIR measurements and interpretations acquired with the ExoMars MicrOmega spare instrument. MicrOmega is a NIR hyperspectral microscope, mounted in the analytical laboratory of the ExoMars rover Rosalind Franklin. All PTAL samples have been observed at least once with MicrOmega using a dedicated setup. For all PTAL samples data description and interpretation are presented. For some chosen examples, RGB images and spectra are presented a well. A comparison with characterizations by NIR and Raman spectrometry is discussed for some of the samples. In particular, the spectral imaging capacity of MicrOmega allows detections of mineral components and potential organic molecules that were not possible with other one-spot techniques. Additionally, it enables to estimate heterogeneities in the spatial distribution of various mineral species. The MicrOmega/PTAL data shall support the future observations and analyses performed by MicrOmega/Rosalind Franklin instrument.

## 1. Introduction

Near-InfraRed (NIR) reflectance spectroscopy has been widely used on Solar System exploration missions in the last 20 to 30 years to characterize the mineralogy and ices on planetary surfaces and on small bodies. In particular, the exploration of Mars has made major advances through the results of the NIR hyperspectral imagers OMEGA aboard Mars Express and CRISM aboard Mars Reconnaissance Orbiter (respectively Observatoire pour la Minéralogie, l'Eau, la Glace et l'Activité, Bibring et al., 2005; and Compact Reconnaissance Imaging Spectrometer for Mars, Murchie et al., 2007). In particular, detection and mapping of several classes of aqueous minerals have revealed the aqueous history of the surface and subsurface of the planet (Gendrin et al., 2005; Poulet et al., 2005; Ehlmann et al., 2008; Milliken et al., 2008; Carter et al., 2013). These spectroscopic observation have driven the selection of the landing sites for the approaching missions that aim to perform in-situ analyses on the surface of Mars, like Mars2020 and ExoMars 2022 (Grant et al., 2018; Loizeau et al., 2019), both targeted to study aqueous evolution and biosignature preservation potential in ancient Mars (e.g. Vago et al., 2017). Aqueous minerals like phyllosilicates, sulfates, hydrated silica or carbonates are of particular interest for these astrobiology missions, as they formed in past environments where stable water was present and they also have high potential for organic molecules preservation (e.g. Farmer and des Marais, 1999). In addition to identifying aqueous minerals, NIR spectroscopy can also identify some organic molecules (for example, Pilorget & Bibring, 2013).

Because of these capacities of NIR spectroscopy with respect to astrobiology missions, instruments dedicated to or capable of NIR spectroscopy are widely used onboard the next in-situ missions to Mars: SuperCam onboard Mars2020, and ISEM (Infrared Spectrometer for ExoMars), Ma-MISS (Mars Multispectral Imager for Subsurface Studies) and MicrOmega onboard ExoMars 2022 will provide spectra of the surface and subsurface and hyperspectral images of rock samples. The spectral domain varies from instrument to instrument with range 1.3-2.6 µm for SuperCam (Wiens et al., 2016), range 1.15-3.3 µm for ISEM (Korablev et al., 2017), range 0.4-2.2 µm for Ma-MISS (De Sanctis et al., 2017), and range 0.99-3.6 µm (and 4 channels in the visible) for MicrOmega (Bibring et al., 2017).

Those missions to the surface of Mars will also benefit from having instruments capable to utilize other techniques like Raman and LIBS spectroscopy. While the SuperCam instrument will combine these three techniques (Wiens et al., 2016), the RLS (Raman Laser Spectrometer) instrument (Rull et al., 2017) will analyze the same samples as MicrOmega during the ExoMars 2022 surface mission. Each technique and instrument is sensitive to different mineral phases or different compositional characteristics. Thus, a coordinated interpretation of results obtained by the different techniques will synergize the identification of the rock composition and of potential organic molecules. This is crucial to enhance the scientific return of these missions.

The Planetary Terrestrial Analogues Library (PTAL) project (Werner et al., 2018; www.PTAL.eu) aims at helping this coordinated work by providing a large set of natural samples and characterizing them with the exact same techniques and/or instruments as on the Mars in situ missions Mars2020 or ExoMars 2022. More details can be found in the first paper of the NIR characterization series that was dedicated to a point spectrometer observation campaign of PTAL crushed-powder samples (Lantz et al., 2020). Full set of samples was characterized by Raman too, and the results have been presented by Veneranda et al. (2019).

The instrument used in this study is a spare of the flight model of MicrOmega (MicrOmega FS) developed for the ExoMars 2022 mission. The main flight model is now mounted on the Rosalind Franklin rover to be launched in year 2022. To take full advantage of the imaging capacity of MicrOmega, bulk rock samples were analyzed, preserving the context of the grains and matrix of the rock, although in the Rosaling Franklin rover, samples will be crushed before MicrOmega analysis, with a significant portion of grains several 100's µm in size (Redlich et al., 2018). Those microscope spectral images enable the identification of mineral species present in minor amount. Furthermore, capability to map the mineral species on the rock surface, over an area of 5 mm x 5 mm, (Field of View –FoV– of MicrOmega FS) provides an estimate of the surficial amount of the different detected mineral species.

After a short description of the PTAL sampling regions, we describe the MicrOmega instrument used in this study, the observation method and the process of data analysis. Then mineralogical characterization is detailed for all PTAL samples as result tables, while some selected samples illustrate the different detections made region by region, comparing shortly this characterization with previously published PTAL studies with an FTIR point spectrometer and Raman spectroscopy. Finally, general observations are made about the results of the mineral characterization of the PTAL samples with MicrOmega.

## 2. The PTAL samples

### 2.1. Provenance of the samples

The PTAL sample collection consists of 99 rocks and sand samples collected from various formation contexts that have been observed or suspected in Mars: lava flows, tuffs, volcanic breccia, hydrothermal environments, solfatara precipitates, sandstones, impact melts and impacted rocks, fracture fills, and with various stages of aqueous alteration in depth or at the surface (Werner et al. 2018). Those are the same samples that were analyzed as crushed powders in Lantz et al. (2020), except for the samples from Lonar crater (India) and Otago (New Zealand). This last location was added only recently to the PTAL collection to include better analogues to the Oxia Planum region, selected as the landing site for the ExoMars 2022 mission (Krzesińska et al., 2021). A more detailed description of all the PTAL sample collection is given in Dypvik et al. (2021). Table 1 (adapted from Lantz et al., 2020) briefly lists the PTAL sample collection.

| *Site name/Rock type* | *Location* | *Sample number-Surface state **(lithology)*** | *Section* |
|---|---|---|---|
| Dry Valleys | Antarctica | DV16-0001 **(gabbro)** | 4.1 |
| Rum | Scotland | RU16-0001 **(ferropicrite)** | 4.1 |
| impact rocks | *various origins* | GN16-0001, VR16-0021 **(impact melt);** WH16-0005, -0014 **(suevite)** | 4.2 |
| impact breccias | Brazil | VA16-0001, VO16-0002 **(polymict breccia);** VO16-0001 **(volcanic breccia)**; VO16-0003 **(Serra basalt)** | 4.3 |
| Jaroso Ravine | Spain | JA08-501 **(jarosite)**; -502 **(quartz)**; -503 **(fracture fill in gneiss)** | 4.4 |
| Rio Tinto | | RT03-501 **(pegmatite w/ quartz)**; -502 **(clay sulphates)**; -503 **(pyrite and sulphates?)** | 4.4 |
| Oslo Rift | Norway | BR16-0001, -0002, UL16-0001 **(gabbro)** | 4.5 |
| Leka Island | | LE16-0001, -0003, -0012, **(harzburgite)**, -0016 **(harzburgite?)**; -0002, -0004, -0005, -0010, -0017 **(dunite)**; -0006, -0007 **(chromite)**; -0008 **(dunite/harzburgite)**; -0009, -0011, -0014 **(gabbroic layer)**; -0013 **(pillow lava)**; -0015 **(serpentine conglomerate)** | 4.7 |
| Reykjanesfólkvangur | Iceland | IS16-0001, -0002, -0003, -0004, -0005, -0013 (**ferropicrite**), -0015, -0016 **(ferropicrite?)**; -0006, -0007, -0008 **(tholeitic lava)**; -0009 **(tholeitic sandstone)**; -0010, -0011, -0012, -0014 **(solfatara precipitation)** | 4.6 |
| John Day Formation | USA | JD16-0001 **(andisol)**; -0002 **(sandstone)**; -0003 **(entisol, weathered tuff)**; 0004 **(altered basalt)**; -0005 , -0007 **(unaltered basalt)**; -0006 **(alfisol, weathered basalt)**; -0008 **(alfisol, weathered tuffaceous siltstone)**; -0009 **(unweathered siltstone)**; -0010, -0012 **(oxisol on rhyolite)**; -0011, -0013 **(rhyolite)**; -0014 to -0019 **(weathered rhyolite)**; -0020, - | 4.8 |
| | | 0021 **(weathered alkali olivine basalt)**; -0022 **(weathered andesite)**; -0023, -0024 **(partly weathered andesite)** | |
| Gran Canaria | Canary Islands | AG16-0001 **(basanite)**; BT16-0001 **(sandstone)**; BT16-0002 **(hyaloclastite)**; CB16-0001 **(pumice/basanite)**, FA16-0001, -0002, -0003 **(altered phonolite)**; RN16-0001 **(olivine/pyroxene alkaline lava)**; TO16-0001 **(tephriphonolite)** | 4.9 |
| Tenerife | | AD16-0001, MR16-0002 **(phonolite)**; AM16-0001, -0002 **(sandstone)**; MR16-0001 **(basanite)**; TF16-0002, -0028, -0059, -0066 **(altered phonolite)** | 4.9 |
| Otago | New Zealand | OT-0001 to -0005 (**conglomerates**) | 4.10 |
| Lonar Crater | India | LO-0001 (**basalt**); LO-0002 and -0003 (**proximal ejecta**) | 4.11 |

*Table 1. Identities and provenance of PTAL samples analyzed in this paper. See the text in dedicated sections for more details. Lithology information with a question mark refers to samples « possibly made of ». This table completes the similar table in Lantz et al. (2020).*

### 2.2. Sample preparation for MicrOmega

Constraints linked to the laboratory setup of MicrOmega FS (detailed in Loizeau et al., 2020) lead to a need to some sample preparation. The short distance between the instrument and the sample and the limited depth of focus of ~0.1mm require a relatively flat top surface of the sample. The cooling system through a cold platform where the sample is set requires a bottom surface of the sample as flat as possible to maximize the contact with the cold platform, and a sample relatively thin to ensure good cooling from bottom to top.

Hence samples from the PTAL collection were sub-sampled and few-mm-thick sections were prepared from each sample. Most of bulk samples were cut with a saw; while for some samples, pieces were collected and then abraded to create thinner fragments with a flat bottom surface. One sample was a sand sample. Eight samples were too small or too fragile to be sawed or sub-sampled for the specific MicrOmega analysis; in their case, only the crushed-powders previously prepared were characterized, and not the bulk rock.

Among sub-samples prepared for MicrOmega studies, the top surface differs from one sub-sample to another (Figure 1). Some sub-samples present past surface weathering or coating, denomited as “natural” surfaces. Other sub-samples have a recently broken surface, either naturally or at the moment of sampling. Finally some sub-samples have been double sawed and hence have both surfaces flat.

Time of operation with MicrOmega FS being limited to preserve the components of this flight spare, and each MicrOmega observation spending about 30 min, it was not possible to perform mosaics of the complete surface of each sample. To be as representative as possible of the bulk composition of the rock, it was decided to avoid when possible exposed surfaces with coating and weathering products. Also, some saw-cut surfaces can have a polish-like aspect that can create unwanted direct reflections of the NIR light with MicrOmega FS, so it was decided to avoid when possible saw-cut or abraded surfaces. Hence, characterization of broken surfaces was preferred during operations. A artcode was applied to name those cut sub-samples by concatenating a letter after the name of the sample: -B (Broken) indicates that the largest top surface of the sub-sample is naturally broken and suffered no evident weathering, -N (Natural) indicates that the top surface experienced significant weathering (alteration or coating), -C (Cut) indicates that the top surface was sawed), -P (Powder) indicates that we analysed the crushed powder, -S (sand) indicates that the sample was sand and not a bulk rock.

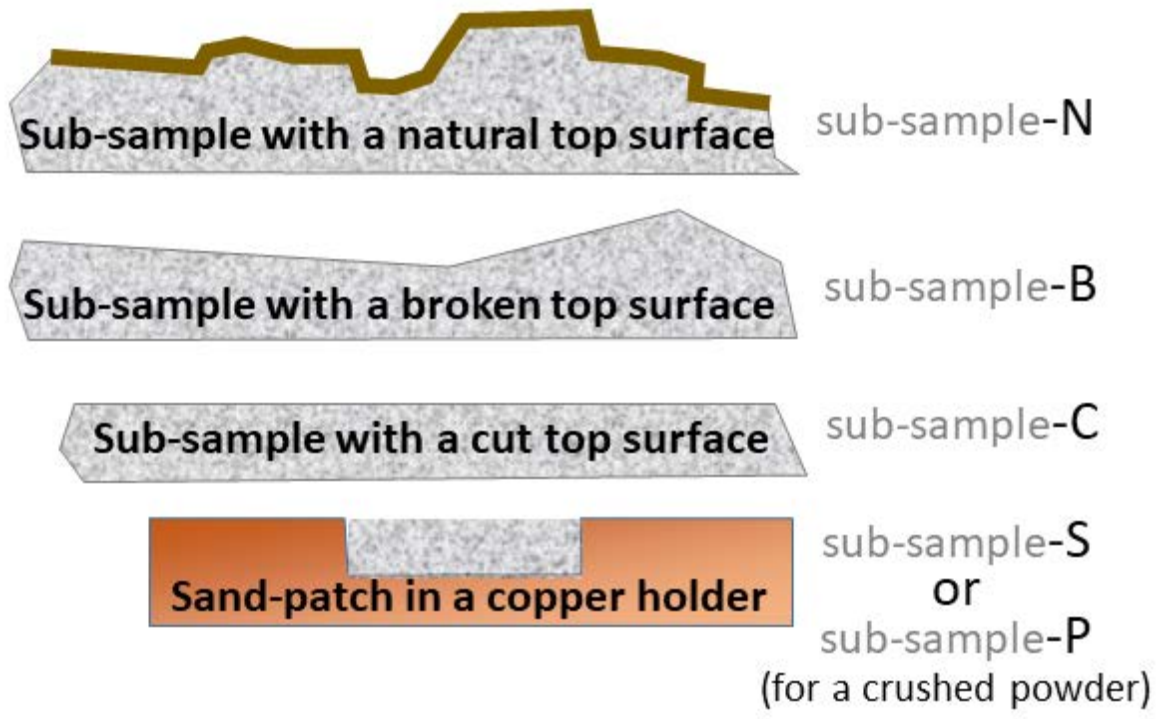

*Figure 1: Different types of sub-samples available for spectral characterization with MicrOmega. Sub-samples are coded with -N, -B, -C, -S or -P to indicate the major state of the surface that was observed.*

# 3. Observation campaign and data analyses method

## 3.1. The MicrOmega instrument

The MicrOmega instrument for ExoMars 2022 (Bibring et al., 2017b) is a microscope acquiring images with pixels of ~20 x 20 µm² over a 256 x 256 pixels field of view (~5 x 5 mm²). An AOTF (Acousto-Optic Tunable Filter) enables to illuminate the FoV with monochromatic light in the NIR range from ~0.99 to ~3.60 µm with a spectral resolution of 20 $cm^{-1}$ (equivalent in wavelength to 2 nm at 1 µm and to 26 nm at 3.6 µm). The cooled detector acquires the reflected light at each of the ~300 wavelengths. The NIR observations using the AOTF are completed with images obtained by illuminating the samples with four different LEDs centered on wavelengths at about 595, 643, 770 and 885 nm (Bibring et al., 2017b). A hyperspectral cube is built by taking acquisitions at different wavelengths and then collecting all images together. Each acquisition is made by the following sequence of images:

- an image is acquired while the FoV is illuminated at a given wavelength, for a chosen integration time,
- a "dark" image is acquired with no illumination of the sample, for the same integration time,
- the difference between those two images is calculated,
- the same set of illuminated/dark images can be accumulated 1, 4 or 16 times at the same wavelength.

The use of MicrOmega FS to observe a large number of samples in safe and efficient conditions required the conception of a dedicated set-up detailed in Loizeau et al. (2020). During data acquisition for the PTAL samples (further named PTAL campaign), the instrument is cooled down by contact with a cold interface at -20°C to -30°C, while the samples are also cooled down by contact with a cold plate at ~-25°C. Those cooler temperatures enable longer integration time, and hence higher signal to noise ratio compared to ambient conditions. They are also closer to the conditions during Mars operations (~-30°C within the ALD – Analytical Laboratory Drawer – where MicrOmega in installed).

## 3.2. MicrOmega observation of the samples

The MicrOmega field of view (FoV) is 5 × 5 mm². Each sample was first inspected visually with an optical microscope to find the most representative area of the sub-sample surface. The sub-samples were then introduced on a cold platform and after temperature equilibrium, were presented in the FoV of MicrOmega for data acquisition (See Loizeau et al., 2020 for more detailed observation preparation). Instrument temperature was approximately constant and integration time was adapted with respect to the albedo of the sample.

During the observation campaign, reference targets were also observed with MicrOmega for radiometric and spectral calibration (Loizeau et al., 2020). The reference targets were observed every time MicrOmega FS was being used after an interruption of more than a week. Specifically, two calibration targets, named *Infragold©* and *Spectralon© 99%* (from Labsphere) provide radiometric calibration. One calibration target named *Spectralon© mixed wavelength standard*, defines the spectral reference and enables to link precisely the acoustic frequency applied to the AOTF crystal to the wavelengths of the emitted light. Those calibration targets are observed at similar instrument and target temperature conditions to those of the sample observations. Those calibration targets are similar to those used for the calibration campaign for the ExoMars mission.

### 3.3. MicrOmega data analysis

Radiometric and spectral calibrations are first performed to create a reflectance cube with two spatial dimensions and one spectral dimension in micrometers according to Riu et al. (2018). Each image generated by MicrOmega FS contains 64000 pixels (256 × 250 pixels), and a spectra is generated for each pixel, hence each spectral cube contains 64000 spectra. For this reason, a first automatic analysis is needed to support a more advanced, manual analysis of the cube. The purpose of this automatic analysis is to bring out the presence of the most common minerals that are detectable in the NIR. For this, we used spectral indices already developed for previous space imaging spectrometers (OMEGA on Mars Express and CRISM on Mars Reconnaissance Orbiter, e.g. Viviano-Beck et al., 2014), adapted for the MicrOmega dataset (Table 2).
The automatic analysis was performed with IDL (Interactive Data Language) routines. It consists of:

- Production of RGB images of the sample scene from the MicrOmega observation,
- Production of maps of >10 spectral indices corresponding to absorptions bands or combinations of absorption bands,

- Checking for the positive detection of the spectral indices based on the spatial coherence of the potential detections (if they are detected on contiguous pixels and not only on isolated pixels, which could correspond to effects of noise, in a minimum number of pixels above a threshold value of the index),
- Calculation of average spectra of the pixels with positive detection for each spectral index.

Figure 2 and Figure 3 show examples of the summary products produced by the automatic analytical tools. Some spectral parameters are sensitive to the presence of more than one mineral or a mineral class. For example, the “OLIVINE” parameter (Table 2) is sensitive to olivine, as its name indicates, but also to the presence of minerals containing iron oxides like iron-rich smectites or iron oxides (Poulet et al., 2007; Viviano-Beck et al., 2014).

Those spectral maps and average spectra can combine several grains of different mineralogy: for example, the spectral parameter named “HYDRATED MINERALS” (Table 2) can be sensitive to grains of different types of phyllosilicates or sulfates, or some carbonates, so a second step of analysis is needed to determine the mineralogical nature. It is out of scope to show these products in this paper but they shall be available for each sample in the PTAL library. In addition, the detection threshold for those spectral parameters cannot be fixed as a constant value for all observations due to the complexity of the surface, of the illumination conditions and the variable signal to noise ratio. It is an evolving compromise between the need to detect all spectral signatures and to avoid too much false detection, and is adapted with the dataset and type of mineral mixtures. Thresholds in Table 2 are an indication of the limit that has been used in most cases with the PTAL MicrOmega dataset.

We chose to make manual and visual analysis as a second step. Each hyperspectral cube is opened with ENVI, to show a visual RGB rendering of the sample, together with maps from the spectral parameters. Local average spectra from several regions of interest within the sample are displayed to look for spectral end-members within the field of view, and spectra are interpreted by comparison with laboratory reference spectra from USGS (Kokaly et al., 2017) or RELAB (RELAB Spectral Database, Copyright 2014, Brown University, Providence, RI.; All Rights Reserved).

Finally, by comparing RGB images, maps of spectral parameters and detections, an estimate was made of the surficial amount of the detected species over the FoV of MicrOmega FS.

Because this estimate is not precisely quantifiable and in order to avoid too much complexity of the result tables of the many samples, we chose to differentiate detections over >50%, from 5 to 50%, over <5% of the FoV, and the absence of detection.

Following this method, we listed mineral detections for each MicrOmega observation presented in section 4. Some missing wavelength channels are present in the MicrOmega FS spectra in the following figures, because of instrumental or calibration artefacts present in the spectra at those channels.

| *Spectral parameter name* | *Mineral species searched* | *spectral band index* | *Position of spectral band(s)* | *Threshold* |
|---|---|---|---|---|
| OLIVINE | Fe-bearing minerals (olivine, Fe-phyllosilicates…) | 1.3-1.7 m slope | Slope from ~1.28 to ~1.7 µm, with wavelength in µm | 4% |
| LCP | low-calcium pyroxene | 1.0-1.2 µm slope | Slope from ~1.0 µm to 1.2 µm, with wavelength in µm | 15% |
| | | 1.8 µm broad band | Band depth centered 1.8-1.9 µm, from continuum at 1.25-1.33 µm to 2.40-2.55 µm | 2.5% |
| HCP | high-calcium pyroxene | 1.0-1.2 µm slope | Slope from ~1.0 µm to 1.2 µm, with wavelength in µm | 15% |
| | | 2.2 µm broad band | Band depth centered 2.15-2.33 µm, from continuum at 1.60-1.70 µm to 2.60-2.67 µm | 2% |
| HYDRATED MINERALS | OH/$H_2O$ phases | 1.4 µm band | Band depth centered 1.41 µm, from continuum at 1.36-1.38 µm to 1.51-1.53 µm | 1% |
| | | 1.9 µm band | Band depth centered 1.91-1.92 µm, from continuum at 1.84-1.86 µm to 2.04-2.07 µm | 1.5% |
| HYDROXYLATED MINERALS | OH phases | 1.4 µm band | Band depth centered 1.41 µm, from continuum at 1.36-1.38 µm to 1.51-1.53 µm | 1% |
| Al-OH MINERALS | Al-OH phyllosilicates | 2.2 µm band | Band depth centered 2.20-2.21 µm, from continuum at 2.11-2.15 µm to 2.24-2.28 µm | 1.5% |
| KAOLINS | kaolin-group | 1.4 µm band | Band depth centered 1.41 µm, from continuum at 1.36-1.38 µm to 1.51-1.53 µm | 1% |
| | | 2.165 µm band | Band depth centered 2.16-2.17 µm, from continuum at 2.10-2.14 µm to 2.22-2.26 µm | 2.5% |
| | | 2.20 µm band | Band depth centered 2.19-2.21 µm, from continuum at 2.10-2.14 µm to 2.22-2.26 µm | 2.5% |
| OPAL | hydrated silica (opal…) | 1.9 µm band | Band depth centered 1.91-1.92 µm, from continuum at 1.84-1.86 µm to 2.04-2.07 µm | 1.5% |
| | | 2.25 µm broad band | Band depth centered 2.20-2.27 µm, from continuum at 2.12-2.15 µm to 2.34-2.36 µm | 2% |
| Fe/Mg-CLAY MINERALS | Fe-rich smectites | 1.9 µm band | Band depth centered 1.91-1.92 µm, from continuum at 1.84-1.86 µm to 2.04-2.07 µm | 2% |
| | | 2.3 µm band | Band depth centered 2.28-2.32 µm, from continuum at 2.23-2.25 µm to 2.34-2.36 µm | 1% |
| SERPENTINE & CHLORITE | Serpentine, Chlorite, Prehnite | 1.39 µm band | Band depth centered 1.39-1.40 µm, from continuum at 1.37-1.38 µm to 1.42-1.43 µm | 0.5% |
| | | 2.32 µm band | Band depth centered 2.30-2.33 µm, from continuum at 2.17-2.21 µm to 2.36-2.38 µm | 1.4% |
| CARBONATES 1 | carbonates | 2.3 µm band | Band depth centered 2.32-2.34 µm, from continuum at 2.19-2.23 µm to 2.37-2.40 µm | 0.5% |
| | | 2.5 µm band | Band depth centered 2.52-2.53 µm, from continuum at 2.40-2.42 µm to 2.60-2.62 µm | 0.5% |
| CARBONATES 2 | carbonates | 3.4-3.5 µm band | Band depth centered 3.4-3.5 µm, from continuum at 3.27-3.29 µm to 3.54-3.55 µm | 1% |
| GYPSUM | gypsum, bassanite | 1.45 µm band | Band depth centered 1.44-1.46 µm, from continuum at 1.36-1.37 µm to 1.64-1.66 µm | 1% |
| | | 1.7 µm band | Band depth centered 1.75-1.77 µm, from continuum at 1.67-1.68 µm to 1.82-1.84 µm | 0.5% |
| | | 2.4 µm slope | Negative slope from 2.32-2.34 µm to 2.42-2.44 µm, with wavelength in µm | 40% |
| AMPHIBOLE | actinolite | 2.30 µm band | Band depth centered 2.30-2.32 µm, from continuum at 2.22-2.26 µm to 2.33-2.37 µm | 1% |
| | | 2.38 µm band | Band depth centered 2.38-2.40 µm, from continuum at 2.35-2.36 µm to 2.43-2.46 µm | 0.7% |
| EPIDOTE | zoisite | 1.67 µm band | Band depth centered 1.66-1.68 µm, from continuum at 1.49-1.50 µm to 1.80-1.82 µm | 1.5% |
| | | 2.47 µm band | Band depth centered 2.46-2.48 µm, from continuum at 2.41-2.43 µm to 2.52-2.54 µm | 1.3% |
| GLASS | basaltic glass | 1.1-1.9 µm slope | Slope from 1.10-1.13 µm to 1.89-1.92 µm, with wavelength in µm | 25% |
| | | 1.95-2.60 µm slope | Slope from 1.94-1.97 µm to 2.57-2.62 µm, with wavelength in µm | 50% |

*Table 2. List of spectral indices used for automatic MicrOmega analysis. Each spectral parameter is the multiplication of the involved spectral band indices. Detection thresholds are adapted for each spectral band and are only a first indication of the likely presence of mineral species.*

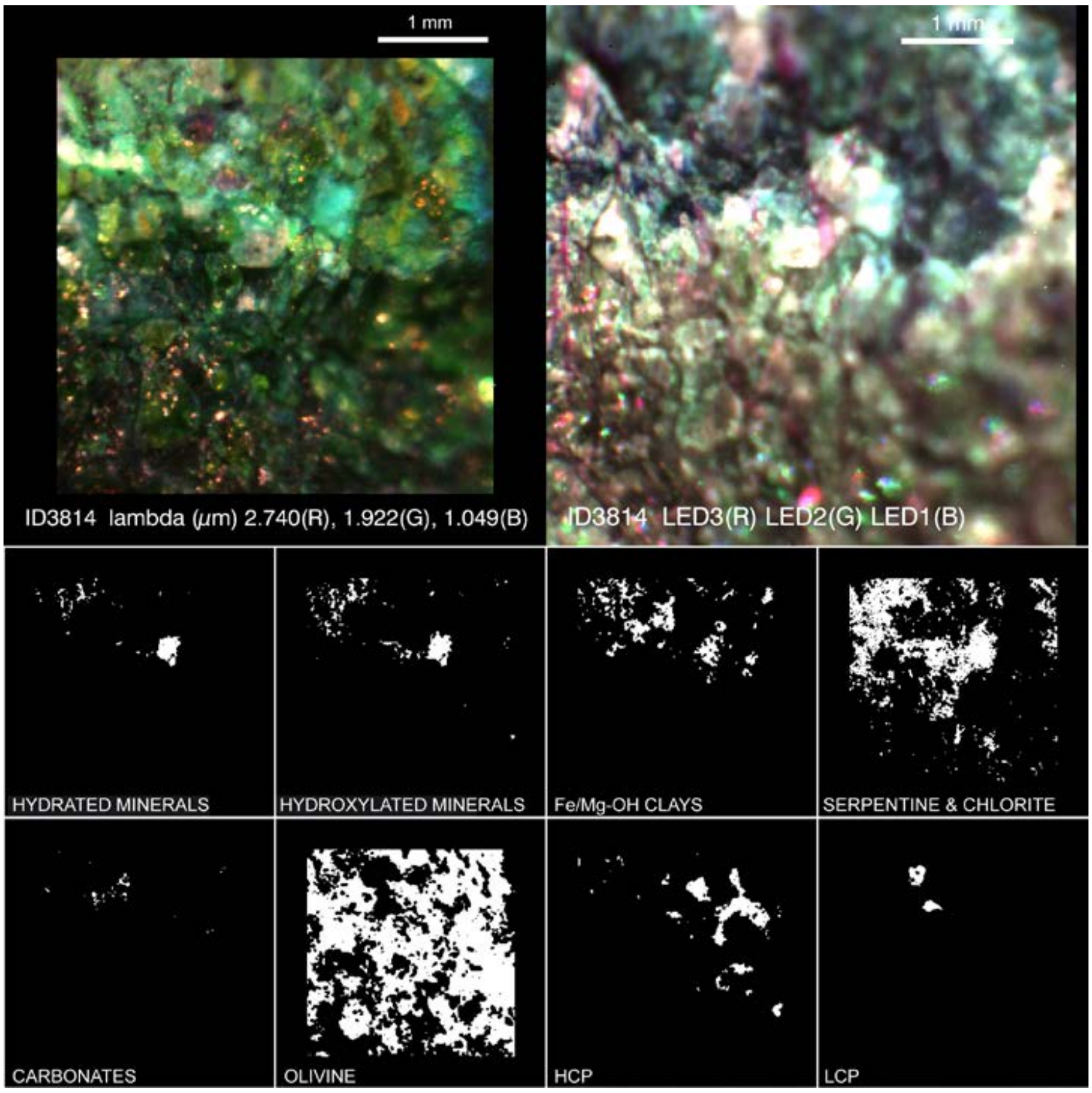

*Figure 2: Example of RGB images for sample RU16-0001, Scotland (in color at the top, with the LED in the visible domain on the right, and the IR light from the AOTF on the left) and spectral parameter maps (in B&W below) produced by the automatic analysis.*

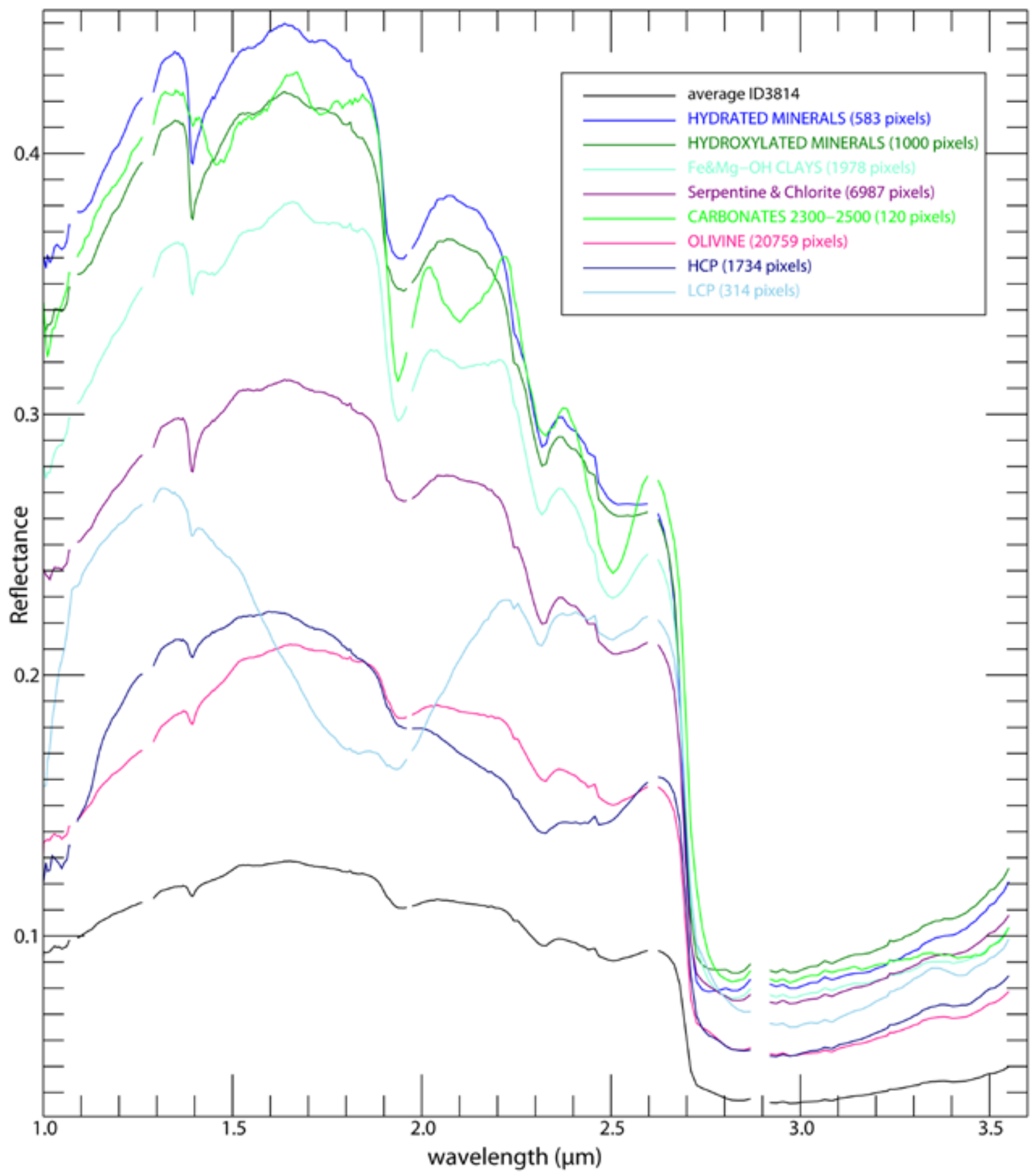

*Figure 3: Example of average spectra for spectral parameters with positive detection (Figure 2) produced by the automatic analyses. These averages of spectra of separated pixels can sometimes increase signal artefacts and are not used for definitive proof of detection but rather as guide for the manual detections.*

## 4. Mineralogical characterization

This section details the mineral detections for each extraction region (Table 1). With about 100 samples, and sometimes more than 5 different spectral endmembers for just one sample, it would not be possible to display the whole spectral variety of all samples in this paper. However, samples have similarities within each sampling region, so we do not display spectra sample by sample but region by region. Even when showing a limited number of selected spectra, the number of figures is still too large for an easily readable article, so we chose to keep only one or two sample example in the main text and to add additional spectral examples in supplementary materials (Annex A). For each region, a table nevertheless lists the detections that are made sample by sample with an estimate of the presence of each detected mineral phase on the surface of studied rocks in the field of view of MicrOmega FS.

In addition, a short comparison is made for each region with previous characterization made on crushed powders of the same samples with a FTIR point spectrometer (Lantz et al., 2020) and with Raman spectroscopy (with two different Raman spectrometers assemblages, Veneranda et al., 2019). Between-technique detection differences are expected. For example the techniques differ in spatial resolution (the FTIR point spectrometer characterized a single spot of ~0.5mm diameter while with Raman, a cumulative number of 30 to 70 focused laser spots were analyzed for each sample). There is also a difference in sensitivity: the FTIR point spectrometer has a wider wavelength domain (0.8 to 4.2 µm vs. ~1.0 to 3.6 µm for MicrOmega FS) allowing an easier detection of oxides and carbonates, and a higher spectral resolution allowing a finer spectral characterization for minerals that are present over the whole FTIR spot; Raman spectroscopy uses a different technique that is sensitive to different minerals, for example quartz and plagioclase are easily detected with Raman spectroscopy and not with NIR reflective spectroscopy, and while clay minerals are more easily distinguishable with the later technique. More specifically, phyllosilicates have weak Raman scattering cross sections compared to the other minerals in most sample and fluorescence from phyllosilicates often masks their relatively weak features. Finally analyses were conducted on different types of samples, namely crushed powders vs. bulk rock surfaces, leading to potential differences depending on the homogeneity of the sample. A more extensive technique comparison is considered for future publication with all techniques used in the PTAL project (XRD, thin section microscopy, NIR spectroscopy and imagery, Raman and LIBS).

Finally, for each sample, spectra of each detected spectral endmember will be available in the PTAL online database accessible through this address: PTAL.eu.

For the sake of conciseness, some samples from different regions are grouped together, like those from section 4.1 below.

### 4.1. Antarctica and Scotland

Samples DV16-0001 and RU16-0001 (Figure 4-1) were collected from altered igneous rocks in the Dry Valleys, Antarctica [Armstrong, 1978] and on Rum Island, Scotland, United Kingdom [Upton et al., 2002] respectively, see Dypvik et al. (submitted) for more details.

Both samples, sampled in far separated continents, show similarities in NIR detection with MicrOmega FS. We detected pyroxenes, both low-Ca pyroxene (LCP) and high-Ca pyroxene (HCP) through large absorption bands centered around 1.8-1.9 µm (LCP) and 2.1 and 2.3 µm (HCP) (see e.g. Klima et al., 2011). HCP spectrum from RU16-0001-B seems more similar to diopside, while HCP spectrum from DV16-0001-B is better fitted with augite (Figure 4-2). Olivine was detected only in RU16-0001-N. Chlorite was detected in both samples (spectra compared to clinochlore in Figure 4-2). Because part of RU16-0001-B has been exposed to the surface for a long time (probably a few years), dry vegetation (potential lichen) has also been detected on this sample (see section 5.2).

Spectral parameter maps (Figure 4-1) give an approximate idea of the proportion of minerals. For example, minerals appearing in purple in the RGB composite of DV16-0001-B are low-Ca pyroxenes crystals and seem to contribute to about 50% of the rock, while high-Ca pyroxenes crystals appear brown in the same RGB composite image and are less present in the rock.

Compared to detection made with a commercial FTIR point spectrometer of crushed powders from the same samples (Lantz et al., 2020), the MicrOmega FS characterization of bulk rock surface expands the mineral detection with the additional detection of high-Ca pyroxene in sample DV16-0001-B and of both low and high-Ca pyroxenes in sample RU16-0001-B. In this last sample, Lantz et al. (2020) identified serpentine instead of chlorite based on small absorption bands at 2.135, 2.44, 2.48, 2.51 and 2.56 µm. These bands are not present in the MicrOmega FS spectra (Figure 4-2) of this sample.

| ***Location*** | ***Sample name*** | ***Sample state*** | ***LCP*** | ***HCP*** | ***olivine*** | ***clinochlore*** | ***dry vegetation*** | ***Additional identified phases in Lantz et al. (2020)*** |
|---|---|---|---|---|---|---|---|---|
| Rum, Scotland | RU16-0001-B | rock | | | L | | | *serpentine* |
| Dry Valleys, Antarctica | DV16-0001-B | rock | L | | | | | *talc or saponite* |

*Table 3. Tables of mineral detections with MicrOmega of the Scotland and Antarctica samples. Black, dark grey, light grey boxes indicate respectively detections over >50%, 5-50%, <5% of the characterized area of the sample, white boxes indicate no detection. An "L" in a grey box indicates that the same mineral was also detected in Lantz et al. (2020) in the powder sample.*

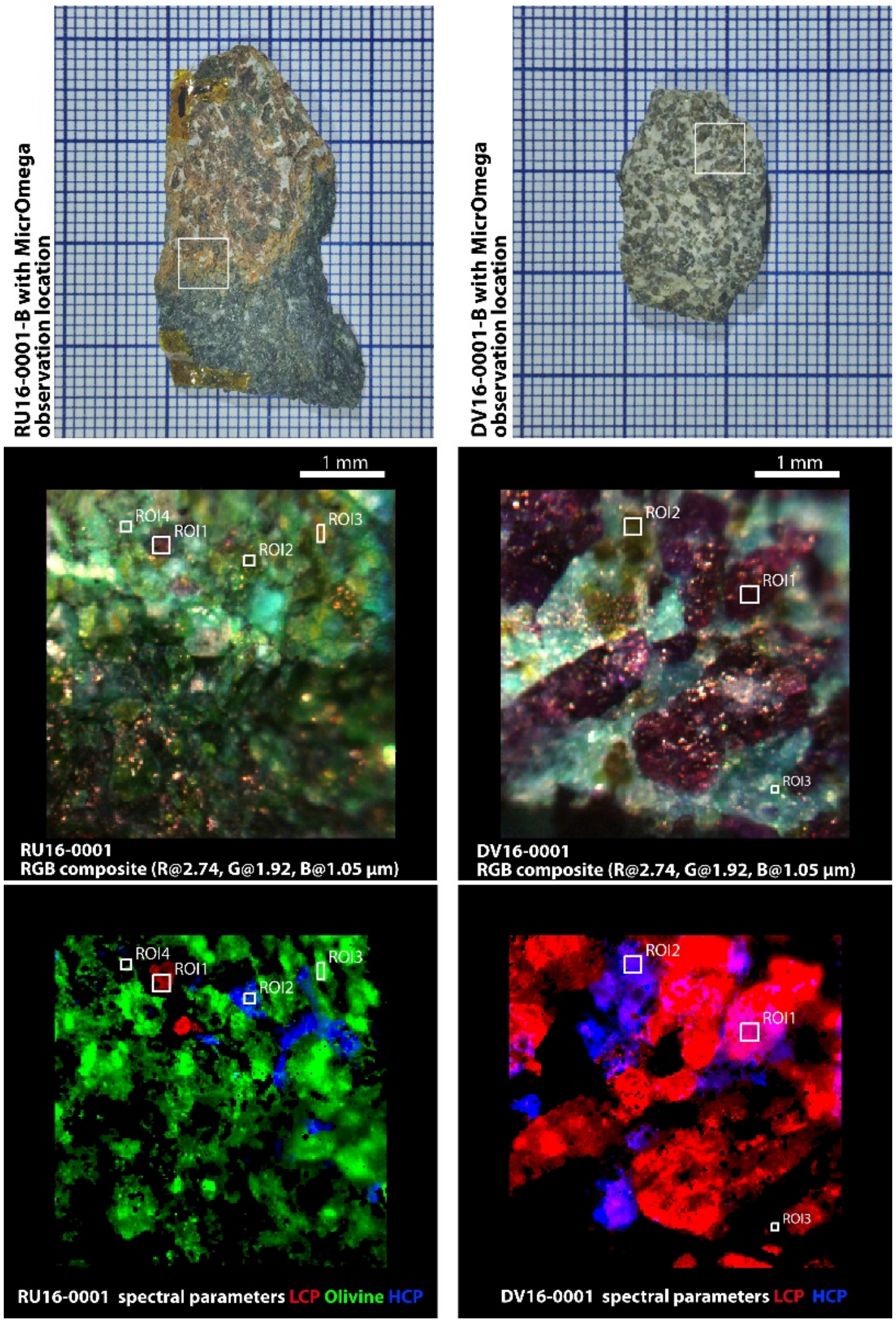

*Figure 4-1. Top to bottom: picture of RU16-0001-B (Scotland) and DV16-0001-B (Antarctica) samples on millimeter paper, RGB composite image of respective MicrOmega observations, RGB composite image of selected spectral parameters. Please note that the maps linked to hydrated minerals are not showed here but hydrated minerals are indeed identified.*

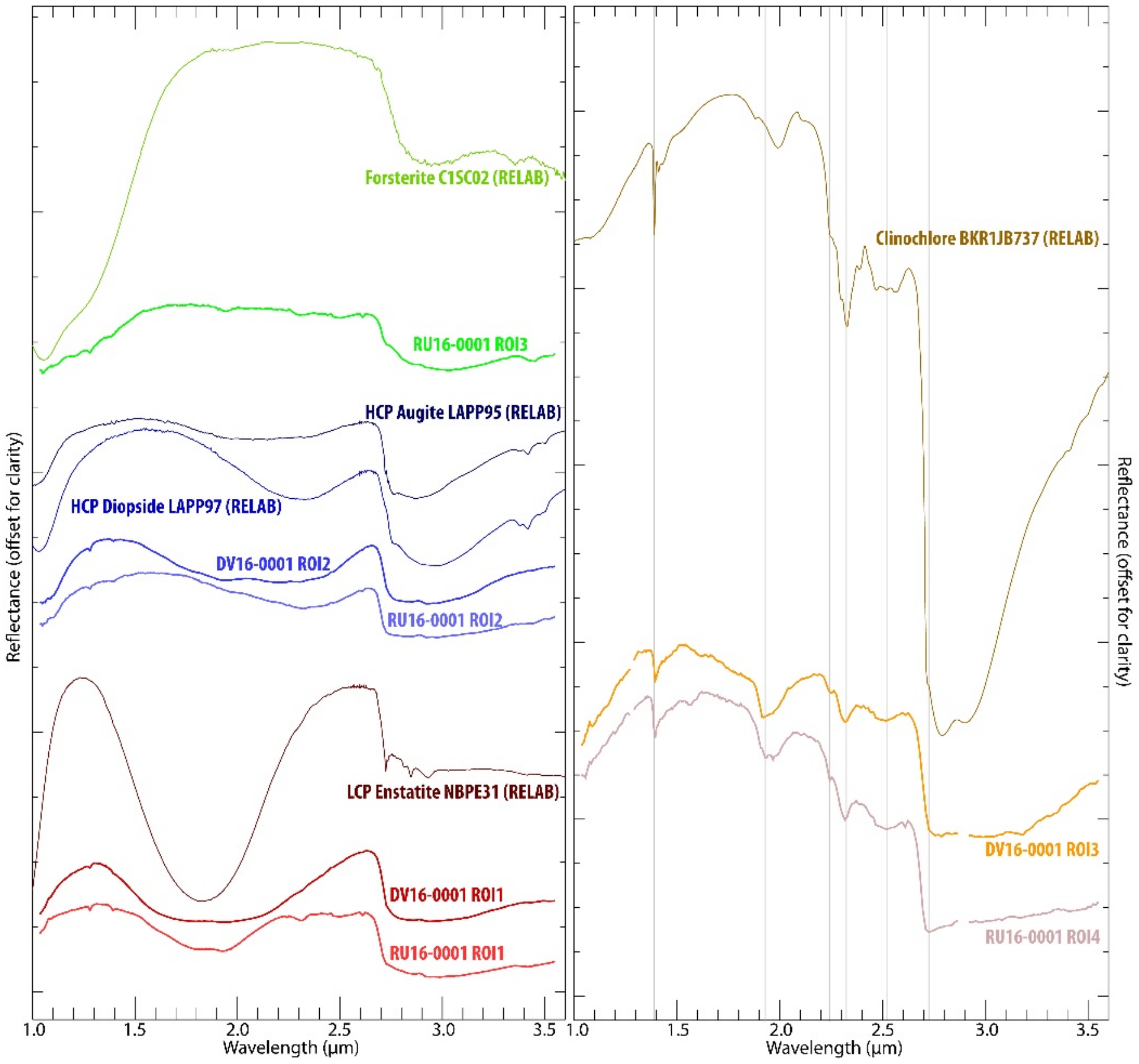

*Figure 4-2. Selection of local averages of MicrOmega spectra compared with reference spectra. Location of Regions of Interest (ROI) in MicrOmega data are indicated in the respective RGB composite images (Figure 4-1).*

## 4.2. Impact melt rocks

These samples have various origins: GN16-0001 was collected in Gardnos, Norway [Kalleson, 2009], VR16-001 in Vredefort, South Africa [Gibson & Reimold, 2008], and WH16-0005 and -0014 come from Chesapeake Bay, USA [Gohn et al., 2009], collected at depth of 1398.26m and 1407.23m respectively.

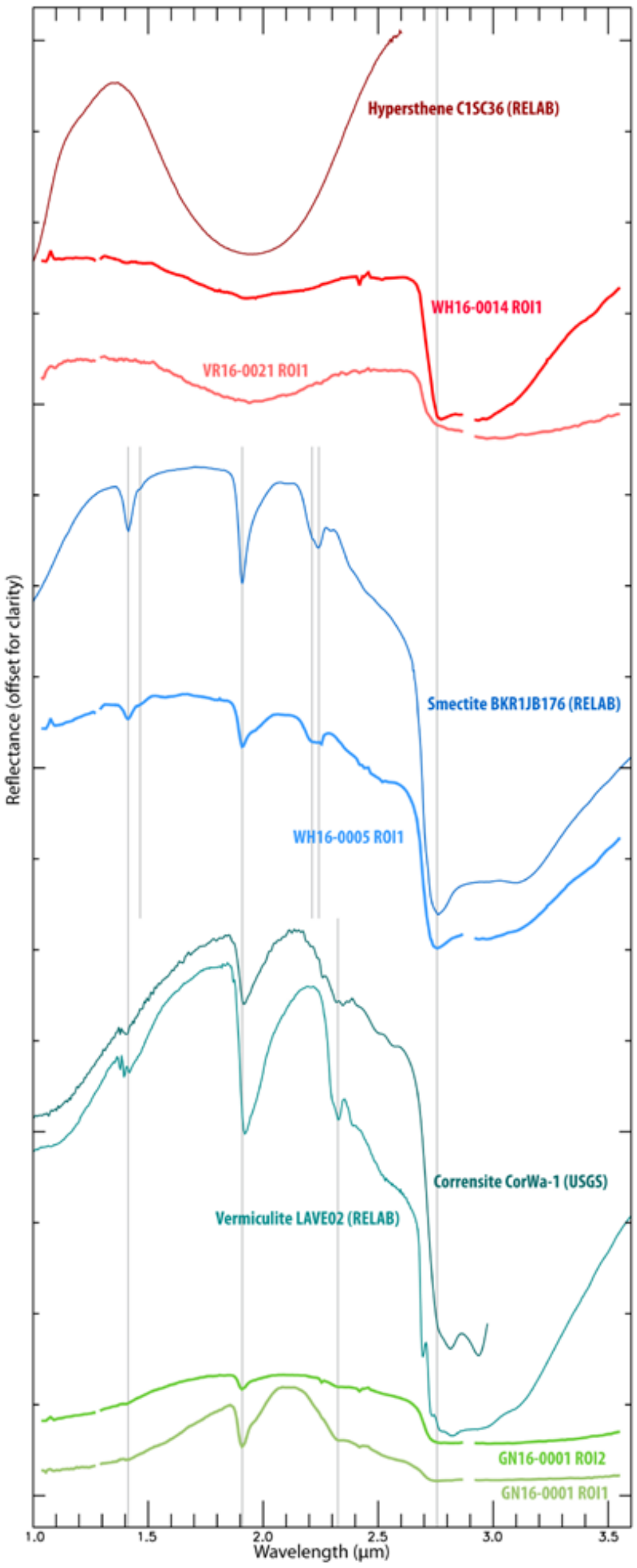

*Figure 5. Selection of local averages of MicrOmega spectra from impact melt rock and crushed powder samples compared with reference spectra. The small spike in some spectra before 1.1 µm is an artefact.*

| ***Location*** | ***Sample name*** | ***Sample state*** | ***LCP*** | ***smectite*** | ***Fe/Mg-phyllosilicate*** | ***Additional identified phases in Lantz et al. (2020)*** |
|---|---|---|---|---|---|---|
| Gardnos, Norway | GN16-0001-B | rock | | | ■ | *Fe oxide* |
| Vredefort, South Africa | VR16-0021-N | rock | L | | | *amphibole* |
| Chesapeake Bay, USA | WH16-0005-P | powder | | ■ | | *carbonate, Fe oxide* |
| Chesapeake Bay, USA | WH16-0014-P | powder | ■ | | | *Al-OH and Si-OH phases* |

*Table 4. Same as table 3 but for the impact melt samples.*

The PTAL samples from Chesapeake bay were too small to saw a dedicated section for the MicrOmega analyses, so only the crushed powder (also used for XRD, Raman and FTIR analyses) were used in this study.

The sample from the Gardnos impact and one sample from the Chesapeake bay deep drill (WH16-0005-B) exhibit clear signature of hydrated minerals (*Figure 5*, absorption bands at 1.41, 1.91). The positive slope from 1.0 to 1.9 µm, the weak 1.4 µm band, and the deep negative slope from 2.2 to 2.3 µm in the spectra from GN16-0001-B may reveal a mixed-layer phyllosilicate (smectite-mica or smectite-chlorite, here compared to a vermiculite and a corrensite) (e.g. Michalski et al., 2015). The large band from 2.20 to 2.24 in the spectra from WH16-0005-P suggests a smectite bearing a mixture of cations (dominantly Al, with additional Fe, Mg…). Samples VR16-0021-N and WH16-0014-P both show the presence of a similar low-Ca pyroxene with a large band centered around 1.9 µm, and a positive slope below 1.2 µm more pronounced in the sample from Vredefort.

Compared to detection made with a commercial FTIR point spectrometer of crushed powders from the same samples (Lantz et al., 2020), the MicrOmega FS characterization of bulk rock surface provides additional information to some mineral identified in those samples. The Fe/Mg-phyllosilicate identified in this study in the Gardnos impact sample was identified by Lantz et al. (2020) as a trace of hydrated phase, as the slope from 2.2 to 2.3 µm was not identified in the more averaged spectrum of the point spectrometer. Low-Ca pyroxene is also identified in WH16-0014-P while it was not with the point spectrometer. On the other hand, carbonate identified in WH16-0005 by Lantz et al. (2020) only with a signature at 3.98 µm cannot be identified with MicrOmega FS in its spectral window 0.99-3.6 µm.

Raman spectroscopy characterization of the same samples (Veneranda et al., 2019) also identified pyroxene in the Vredefort sample, but not in the WH16-0014 sample. Raman spectroscopy could not identify smectite or phyllosilicate in any of these samples, but this technique is not very sensitive to these mineral species.

### 4.3. Brazil: breccia from impact in basaltic formations

Sample VA16-0001 was collected at Vista Alegre, Paraná [Crosta et al., 2010] while samples VO16-0001 to -0003 come from Vargeão Dome, Santa Catarina [Crosta et al., 2012], both locations in Brazil. Those samples come from impact breccia formed through the impact that created both craters.

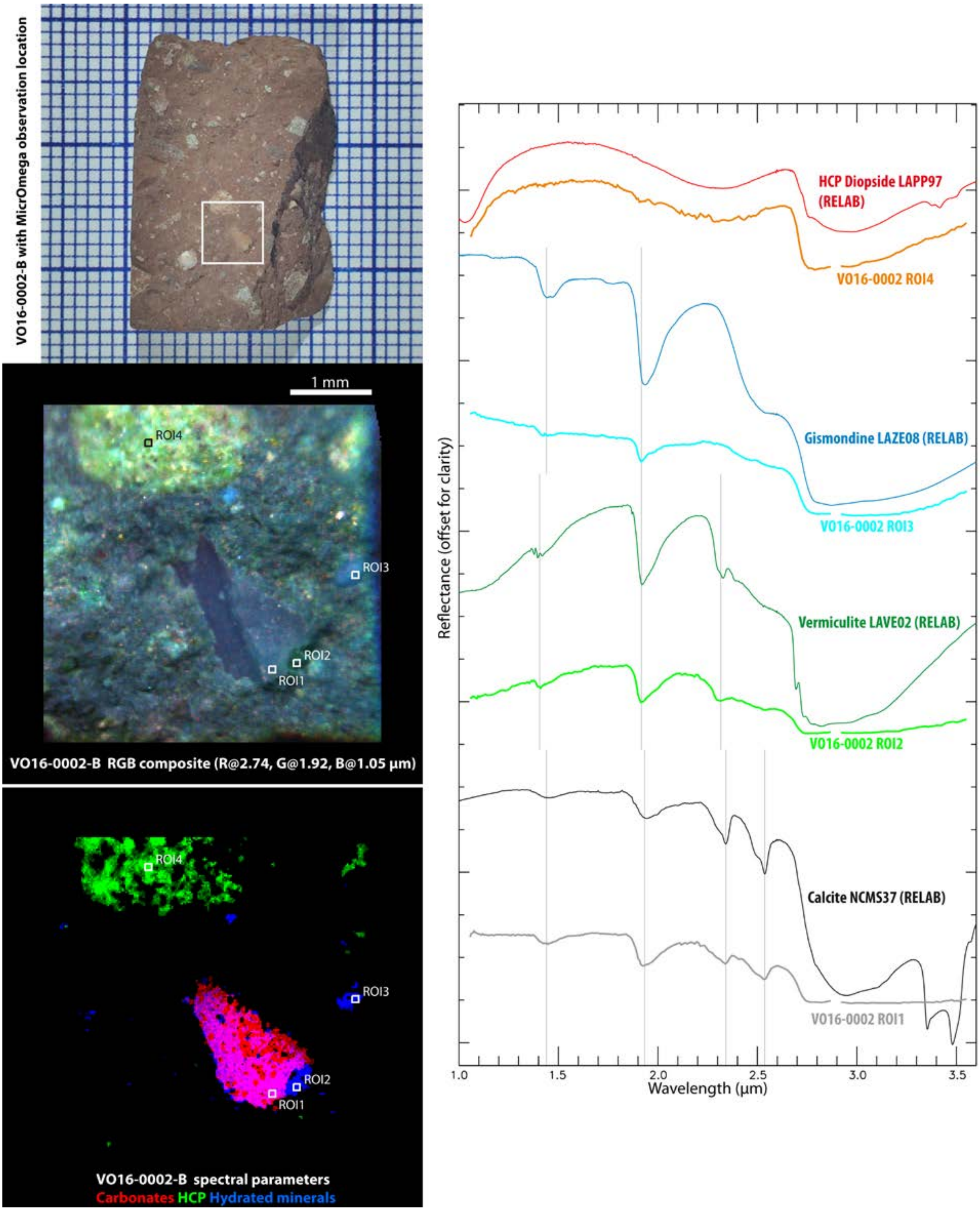

*Figure 6. From left to right and top to bottom: picture of VO16-0001-B sample on millimeter paper, RGB composite image of MicrOmega observation, RGB composite image of selected spectral parameters, and selection of local averages of MicrOmega spectra compared with reference spectra. Location of Regions of Interest (ROI) in MicrOmega data are indicated in the RGB composite image.*

Only one example is illustrated in this section (VO16-0002-B, Figure 6) because of the four samples from Brazil, it shows a particularly high spectral variety with well-defined spatial microscopic structures. The MicrOmega spectral cube indeed indicates the presence of pyroxene (here compared to diopside) as a ~3 mm large bright grain, carbonate (here compared to calcite) as a triangular ~1 mm large dark grain, zeolite (here compared to gismondine) and a Fe/Mg-rich phyllosilicate (here compared to vermiculite) as sub-mm grains (Figure 6).
All Brazil impact breccia samples show pyroxenes and hydrated minerals, zeolites or phyllosilicate or carbonates or oxide hydroxide.

Compared to detection made with a commercial FTIR point spectrometer of crushed powders from the same samples (Lantz et al., 2020), the MicrOmega FS characterization of bulk rock surface completes the list of detected minerals: pyroxene was not detected at all in VA16-0001 and VO16-0002, low-Ca pyroxene was not detected in VO16-0001, and in VO16-0003, the detection was less clear for the Fe oxide and the phyllosilicate.

Interestingly, the crushed powder from this sample has also been observed with MicrOmega FS (Figure 7). The average spectrum of this observation is very similar to the point spectrometer observation from Lantz et al. (2020). But the fine powder (largest particules >100 µm but average size <20 µm, the size of one MicrOmega FS pixel) enables the identification of less minerals than the bulk rock observation (Figure 6): high-Ca pyroxene and hydrated phyllosilicates (possibly a zeolite) could be identified, but there were no spectral band at 2.3 µm or 2.5 µm for Fe/Mg-bearing phyllosilicates or carbonates. As this sample has a high heterogeneity, some mineral species may be present on the bulk rock surface but absent in the crushed powder.

Raman spectroscopy characterization of the same samples (Veneranda et al., 2019) also identified pyroxene in most samples, carbonate in VA16-0001 and VO16-0002, and did not identify the different phyllosilicate detected here in the three samples from Vargeão Dome.

| Location | Sample name | Sample state | LCP | HCP | carbonate | Fe/Mg-smectite or vermiculite | zeolite | Fe oxi hydroxide | Al-OH phyllosilicate | unidentified hydrated phase | Additional identified phases in Lantz et al. (2020) |
|---|---|---|---|---|---|---|---|---|---|---|---|
| Vista Alegre | VA16-0001-B | rock | | | L | | | | | L | olivine, Fe/Mg-smectite |
| Vargeão Dome | VO16-0001-B | rock | | L | | L | | | | | |
| Vargeão Dome | VO16-0002-B | powder | | | L | L | | | | | Fe oxide |
| Vargeão Dome | VO16-0003-B | powder | | L | | | | L | | | |

*Table 5. Tables of mineral detections with MicrOmega of the Brazil samples.*

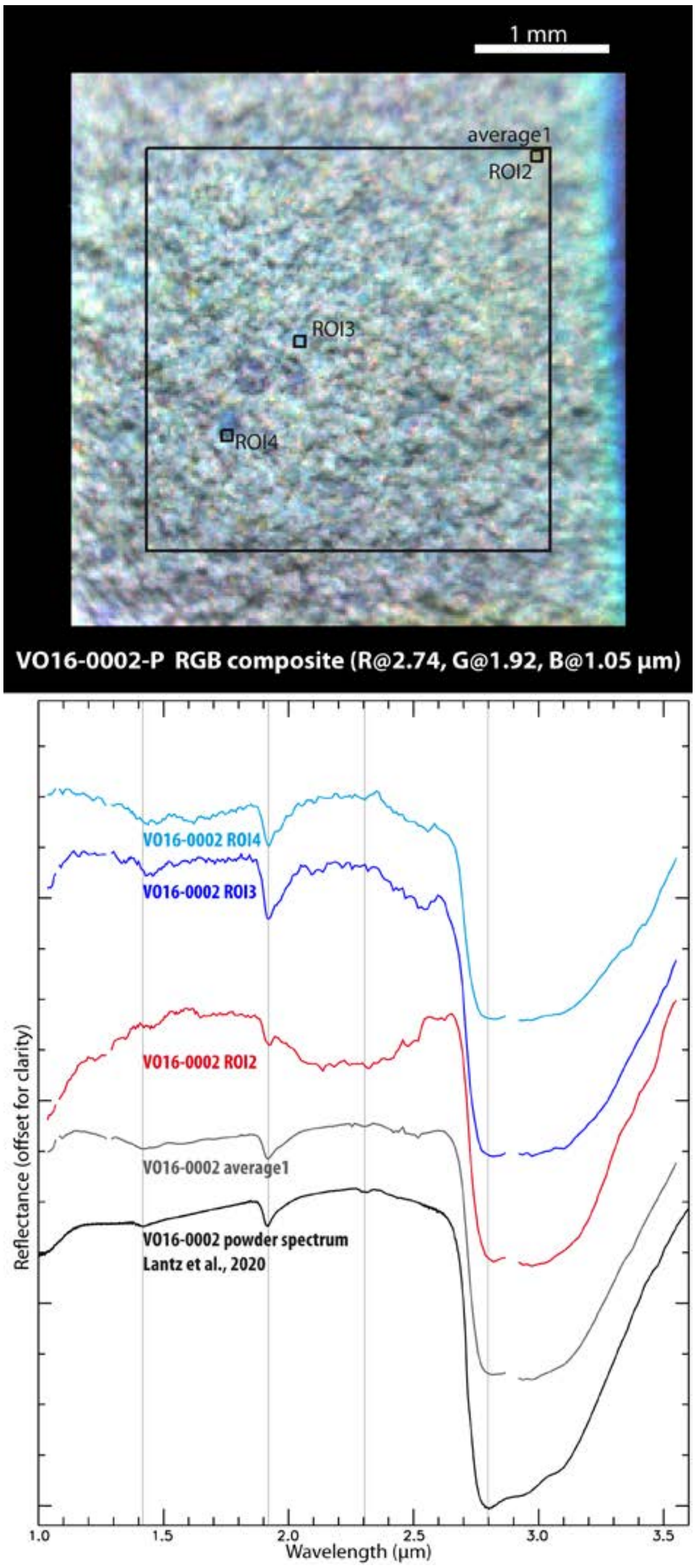

*Figure 7. Observation of sample VO16-0002 (as in the previous figure) but as a crushed-powder sample. From top to bottom: RGB composite image of MicrOmega observation, selection of large and local averages of MicrOmega spectra, and spectrum of the same powder sample acquired with a point spectrometer (Lantz et al., 2020). Location of Regions of Interest (ROI) in MicrOmega data are indicated in the RGB composite image. Local averages are of the same size as in the previous figure (5 pixels by 5 pixels).*

### 4.4. Jaroso Ravine and Rio Tinto

These samples from Spain were collected at Jaroso Ravine (site of Jarosite discovery) for JA08-501, -502, and -503 [Martinez-Frias et al., 2004], and at Rio Tinto for RT03-501, -502, and -503 [Hudson-Edwards et al., 1999].

Most of the samples we could use were too small and fragile to saw a part of it for bulk analysis, so except for RT03-501, MicrOmega characterization was made on crushed powder samples only (grain size < 100 μm).

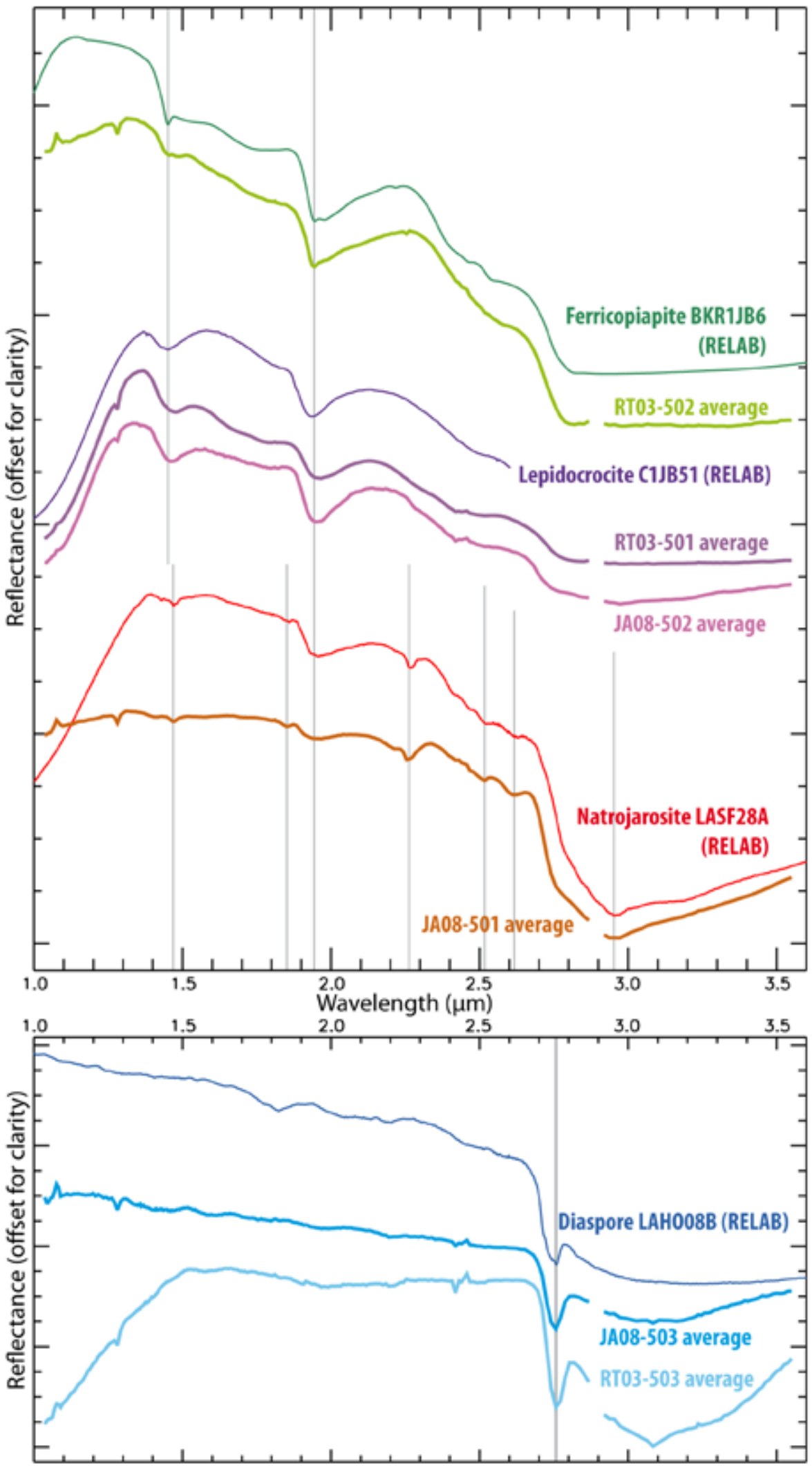

*Figure 8. Selection of local averages of MicrOmega spectra from Jaroso Ravine and Rio Tinto rock and crushed powder samples compared with reference spectra.*

Fine crushed powder made it more difficult to separate spatially mineral grains with MicrOmega images, so we present here average spectra over most of the field of view of MicrOmega FS, and not on local ROI averages as for other samples in this study.

These samples show the presence of sulfates (jarosite, copiapite) and oxide-hydroxide (lepidocrocite and diaspore) (Figure 8, Table 6). Similar mineralogy has already been identified in the Rio Tinto formation with reflectance visible-near infrared spectroscopy with detection of copiapite and oxide-hydroxides as main rock components (Kaplan et al, 2016).

| ***Location*** | ***Sample name*** | ***Sample state*** | ***jarosite/ natrojarosite*** | ***Fe oxi hydroxide*** | ***Fe oxide*** | ***Al oxi hydroxide (diaspore)*** | ***sulfate (possible alunite)*** | ***sulfate (possible ferricopiapite)*** | ***Additional identified phases in Lantz et al. (2020)*** |
|---|---|---|---|---|---|---|---|---|---|
| Jaroso Ravine | JA08-501-P | powder | L | | | | | | |
| Jaroso Ravine | JA08-502-P | powder | | L | | | | | *carbonate* |
| Jaroso Ravine | JA08-503-P | powder | | | | L | | | *Fe-oxide* |
| Rio Tinto | RT03-501-B | rock | | L | | | | | |
| Rio Tinto | RT03-502-P | powder | | L | | | | L | |
| Rio Tinto | RT03-503-P | powder | | | L | L | | | *carbonate* |

*Table 6. Tables of mineral detections with MicrOmega of the Spain samples.*

MicrOmega FS characterization in this study and previous FTIR point spectrometer analyses are very well in agreement for these samples as both study were made on fine homogeneous crushed powders for all samples but RT03-501. Only carbonate, identified at 3.98 and 4.0 µm in Lantz et al. (2020) could not be identified, as being outside of the spectral domain of MicrOmega FS.

Raman spectroscopy (Veneranda et al., 2019) also identified jarosite only in sample JA08-501, Fe oxide hydroxide (as goethite) was also identified in JA08-502 and RT03-501 but not in RT03-502 (where MicrOmega FS spectra are more similar to lepidocrocite than goethite), and no oxide was identified in RT03-503 while MicrOmega FS detected both Fe oxide and potential Al oxide hydroxide. Raman detections from Veneranda et al. (2019) were made both with the RLS simulator (exciting laser at 532 nm) and the microRaman (exciting laser at 633 nm). A possible explanation for this lack of detection with Raman may be related to grains too small in the crushed powder that was investigated for this sample. Veneranda et al. (2019) also identified a sulfate (barite) in RT03-501 where this study detects possible alunite.

## 4.5. Oslo rift samples

These samples come from the Norwegian regions of Brattåsen for BR16-0001 and -0002, and Ullernåsen for UL16-0001 [Neumann et al., 1985].

Sample UL16-0001-B from Ullernåsen shows the presence of chlorite and carbonate (Table 7, Annex A Figure 1), while the samples from Brattåsen BR16-0001-B and BR16-0002-B show both high-Ca pyroxene (Annex A Figure 2) and amphibole (spectrally similar to actinolite), and chlorite and Fe/Mg-phyllosilicate (possibly vermiculite) for the first one, and likely iron oxide for the second.

In UL16-0001, chlorite and carbonate were also identified by Lantz et al. (2020) using a FTIR point spectrometer. Interestingly, carbonate was only identified with the double band at 3.85+3.98 µm in the larger average spectrum of the point spectrometer, while targeting grains with MicrOmega FS revealed as well bands at 2.3, 2.5, and the doublet around 3.4 and 3.5 µm (Annex A Figure 1). In both samples from Brattåsen, the point spectrometer could also identify high-Ca pyroxene and amphibole, but did not detect any phyllosilicate (chlorite, smectite or vermiculite) in BR16-0001.

In Raman data, Veneranda et al. (2019) reported pyroxene and carbonate in all three samples, but did not detect amphibole, neither phyllosilicates. They also detected gypsum in UL16-0001 that was not detected on the bulk rock sample in the present work.

| ***Location*** | ***Sample name*** | ***Sample state*** | ***HCP*** | ***amphibole (actinolite)*** | ***chlorite*** | ***iron oxide*** | ***carbonate*** | ***Fe/Mg-smectite or vermiculite*** | ***Additional identified phases in Lantz et al. (2020)*** |
|---|---|---|---|---|---|---|---|---|---|
| Ullernåsen | UL16-0001-B | rock | | | L | | L | | *Fe oxide* |
| Brattåsen | BR16-0001-B | rock | L | L | | | | | *olivine, carbonate* |
| Brattåsen | BR16-0002-B | rock | L | L | | L | | | *carbonate* |

*Table 7. Tables of mineral detections with MicrOmega of the Oslo rift samples.*

We added in Figure 9 a comparison of bulk rock and powder sample observations of BR16-0001, itself compared with the observation of the crushed powder with the point spectrometer in Lantz et al. (2020). The crushed powder shows several larger grains up to >200 µm. The average Micromega FS observation of the crushed powder is well comparable with the point spectrometer observation, although with an expected lower spectral resolution, and a slope difference between 2.8 and 3.0 µm. The deep band at 2.73 µm is present in most spectra of the powder sample, this is likely due to the presence of fine particles (<20 µm) of actinolite in almost all MicrOmega FS pixels. The presence of actinolite is suspected do to the fine band at 1.4 µm, the two bands at 2.30-2.32 and 2.38 µm, and the deep band at 2.73 µm (e.g. Mustard, 1992, see reference spectrum of actinolite in Annex A, Figure 1). Mineral species like actinolite and chlorite seem to be detected in both rock and powder observations, but with weaker signatures for the crushed powder. Fe/Mg-phyllosilicate (possibly vermiculite) has been detected on the bulk rock sample but could not be detected in the powder sample, but a grain with a 2.2 µm band which could be a kaolinite was detected only in the powder sample.

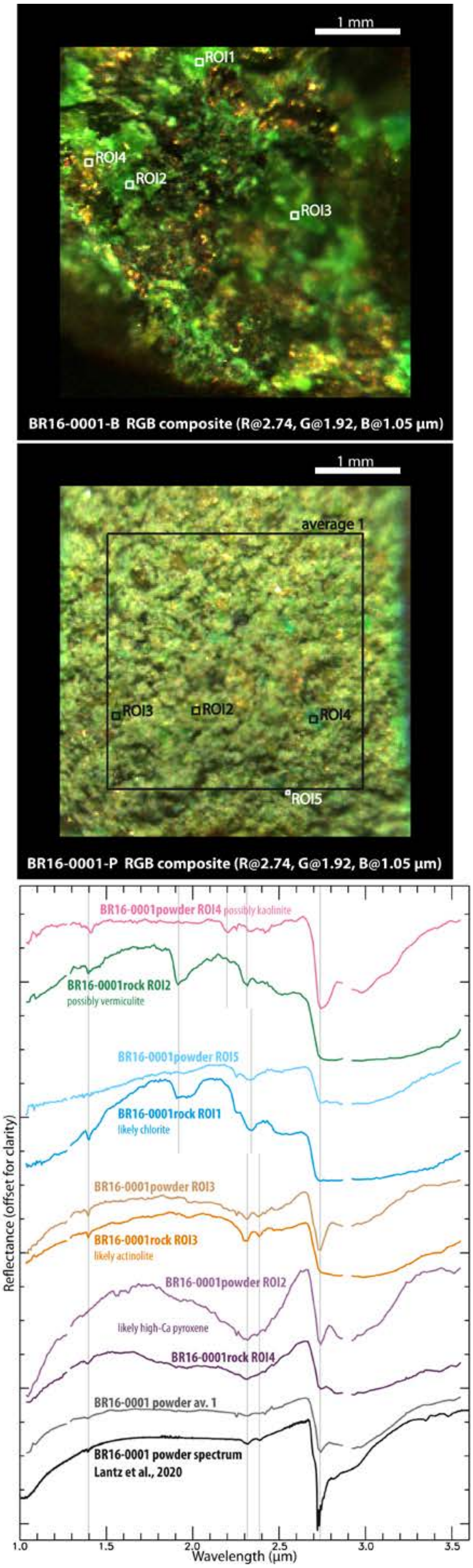

*Figure 9. Micromega observations of sample BR16-0001 both as a bulk rock surface (top) and as a crushed-powder sample (middle). From top to bottom: RGB composite images of MicrOmega observation of the bulk rock and of the crushed powder, selection of local averages of MicrOmega spectra for comparison between the bulk rock and the crushed powder, and average spectrum of the crushed powder observation compared with spectrum of the same powder sample acquired with a point spectrometer over a field of ~1 mm² (Lantz et al., 2020). Location of Regions of Interest (ROI) in MicrOmega data are indicated in the RGB composite image. Local averages are of the same size (5 pixels by 5 pixels) except for ROI5 of the crushed powder observation (3 pixels by 3 pixels).*

| Location | Sample name | Sample state | *HCP* | *olivine* | *basaltic glass* | *Fe oxi hydroxide* | *possible Fe oxide* | *hydrated silica (opal)* | *kaolinite/ halloysite* | *sulfate (possible copiapite)* | *sulfate (polyhydrated)* | *organic phase* | *unidentified hydrated phase* | *possible chloride* | ***Additional identified phases in Lantz et al. (2020)*** |
|---|---|---|---|---|---|---|---|---|---|---|---|---|---|---|---|
| Haleyabunga | IS16-0001-C | rock | L | L | | | | | | | | | | | |
| Haleyabunga | IS16-0002-C | rock | L | | | | | | | | | | | | |
| Lagafell | IS16-0003-B | rock | L | L | | | | | | | | | | | |
| Lagafell | IS16-0004-C | rock | L | L | | | | | | | | | | | |
| Lagafell | IS16-0005-C | rock | L | | | | | | | | | | | | |
| Stapafell | IS16-0006-S | sand | | | L | | | | | | | | L | | |
| Stapafell | IS16-0007-C | rock | | L | | | | | | | | | | | *HCP* |
| Stapafell | IS16-0008-B | rock | | L | | | | | | | | | | | *HCP* |
| Stapafell | IS16-0009-B | rock | | | L | | | | | | | | L | | *Si-OH phase* |
| Seltun | IS16-0010-B | rock | | | | | | L | | | | | | | *Fe-oxide* |
| Seltun | IS16-0011-N | rock | | | | | | | | | | | | | *olivine, kaolin* |
| Seltun | IS16-0012-C | rock | | | | | | L | | | | | | | |
| Haleyabunga | IS16-0013-N | rock | L | L | | | | | | | | | | | |
| Reykjanes | IS16-0014-B | rock | | | | | | | L | | | | | | *Fe oxide* |
| Vatnsheidi | IS16-0015-N | rock | | L | | | | | | | | | | | *HCP* |
| Vatnsheidi | IS16-0016-C | rock | | | | | | | | | | | | | *HCP, Si-OH phase* |

*Table 8. Same as Table 3 but for the Iceland samples.*

### 4.6. Iceland samples

These samples come from the Southern Peninsula region, close to Reykjavík, in the areas of Haleyabunga (IS16-0001, -0002, -0013), Lagafell (IS16-0003, -0004, -0005), Stapafell (IS16-0006, -0007, -0008, -0009), Seltun (IS16-0010, -0011, -0012), Reykjanes (IS16-0014), and Vatnsheidi (IS16-0015, -0016) [Sigmarsson and Steinthorsson, 2007].

The Iceland samples are diverse: ferropicrite basaltic rocks (basalt that is very rich in Mg-rich-olivine, and with a relatively high Fe content, IS16-0001 to -0005, -0013, -0015 and -0016), pillow lavas (IS16-0007 and -0008), sand/sandstone of eroded lavas (IS16-0006 and -0009), and rocks from solfatara (fumaroles emitting sulfurous gases, IS16-0010 to -0012 and -0014).

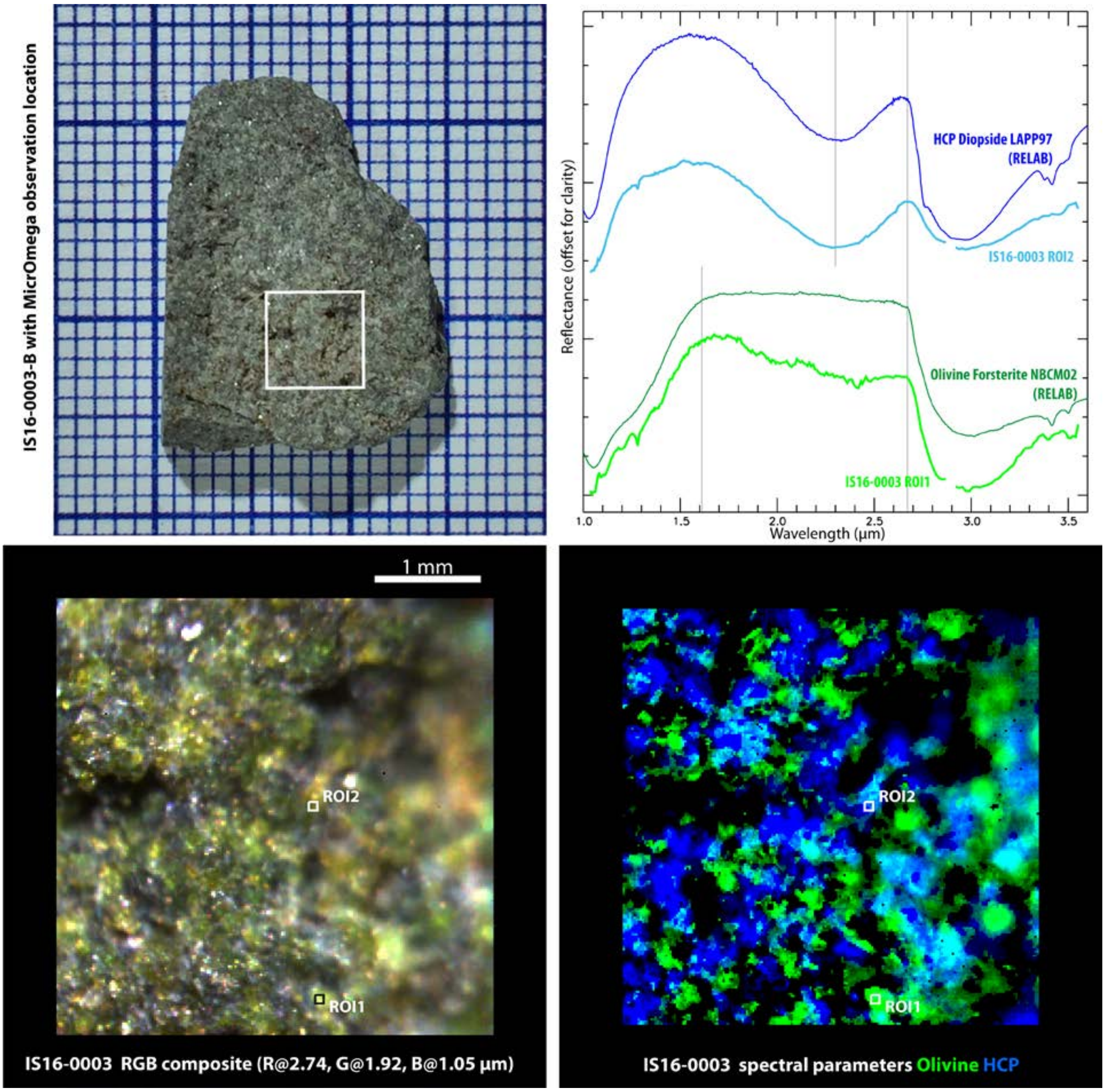

*Figure 10. Same as Figure 6 but for sample IS16-0003-B. As the sample is uneven across the FoV, part of the imaged area is unfocused, creating the blurry area on the right part of the MicrOmega image and map. The small spike around 1.3 µm is an artefact.*

High-Ca pyroxene and olivine are generally present in the pillow lavas and ferropicrite basaltic samples (Table 8, e.g. Figure 10) except for IS16-0016-C that only shows presence of iron oxide in the matrix and oxide-hydroxide within vessicles (Figure 11). This indicates a much higher degree of oxidation for this sample compared to the others. Vessicles (0.5 mm large) in IS16-0015-N reveal the presence of a polyhydrated sulfate, but this could have been deposited recently as the characterized surface of the sample was exposed to the surface before being collected. Similarly, the organic phase detected through a ~3.4 µm band in the vessicles in sample IS16-0005-C looks like small radicles that could have penetrated the vessicles inside the rock before it was sawed.

Sand and sandstone from eroded lava (IS16-0006-S and -0009-B) contain probable basaltic glass grains (Annex A Figure 3), together with olivine and hydrated phases. The presence of a chloride is likely in IS16-0006-S.

As expected, samples from solfatara (IS16-0010-B and -0012-C) differ completely spectrally with the presence of altered phase such as hydrated silica (opal-like), as well as kaolinite and sulfate in the later (Annex A Figure 4).

Pyroxene was also identified with a FTIR point spectrometer by Lantz et al. (2020) in the same samples as this study, except for IS16-0011. Conversely, the signatures are too shallow to conclude on a positive detection with MicrOmega FS for IS16-0007, -0008, -0015 and -0016. Olivine detections between the FTIR and MicrOmega FS are also consistent, except for sand- and sandstone-samples IS16-0006 and -0009 where olivine was only detected with MicrOmega FS in isolated grains (<0.1 mm large). Basaltic glass was also detected in the same sand and sandstone samples, and opal was definitely identified in the same solfatara samples. The strongest difference comes with sample IS16-0016 where Lantz et al. (2020) identified pyroxene and possible hydrated silica, whereas MicrOmega FS spectra point to the presence of Fe-oxides (hematite and oxide hydroxide like lepidocrocite).

Raman spectroscopy analyses (Veneranda et al., 2019) indicate pyroxene and olivine in all samples where MicrOmega FS detected some. Basaltic glass was also identified in the same sand and sandstone samples. Fe-hydroxide was identified in IS16-0016 but as goethite and not lepidocrocite. While MicrOmega FS sees a sulfate in IS16-0012, they identified sulfur. Hydrated silica, kaolinite and sulfates could not be identified with Raman spectroscopy.

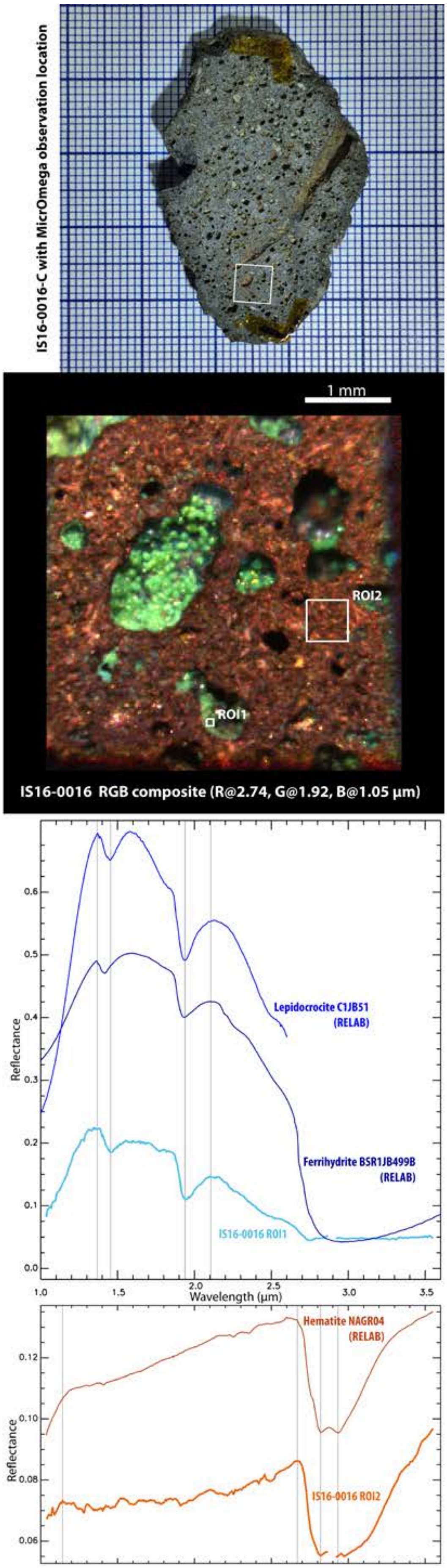

*Figure 11. Same as Figure 6 but for sample IS16-0016-C.*

## 4.7. Leka Samples

All these samples come from Leka Island, Norway. The Leka Ophiolite Complex represents a part of the oceanic lithosphere which has been extensively serpentinized at the ocean floor [Furnes et al., 1988]. A wide range of temperatures and different primary minerals and rocks have created a variety of metamorphic secondary minerals.

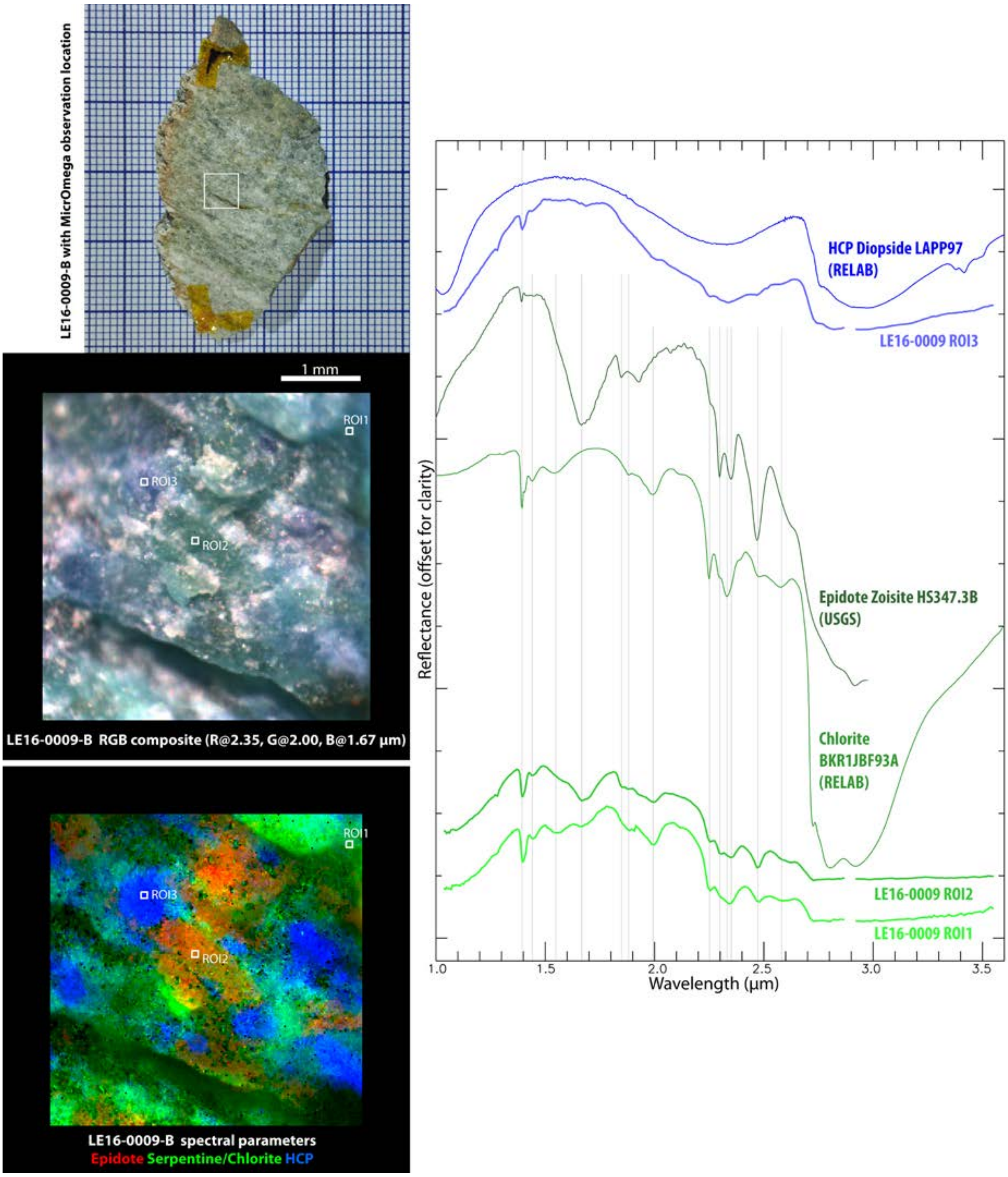

*Figure 12. Same as Figure 6 but for sample LE16-0009-B.*

Leka samples mineralogy is dominated by serpentine (Table 9, Annex A Figure 5), detected in 13 samples out of 17. Serpentine was actually the only mineral species that could be identified with MicrOmega in most samples. But a few samples show more diversity and the presence of chlorite as LE16-0009, -0011, -0014 and -0015. Chlorites can be spectrally distinguished from Mg-rich serpentine (e.g. King and Clark, 1989). For example, spectra from LE16-0009-B (Figure 12) show no serpentine, but chlorite, the epidote zoisite (through the bands at 1.66, 1.85, and 2.48) and high-Ca pyroxene. Chlorite is always detected when serpentine is not detected in the Leka samples (Table 9), and epidote and/or pyroxene is always present in the samples with chlorite (Annex A Figure 6). Carbonate was only detected in one sample (LE16-0014-N) and could come from surface weathering.

When compared with results of the Leka samples from FTIR point spectrometry (Lantz et al., 2020) and Raman spectrometry (Veneranda et al., 2019), there is an excellent consistency for the presence or absence of serpentine. Lantz et al. also identified chlorite in samples LE16-0009, -0011, -0013 and -0014, and epidote in LE16-0009 and -0014. Veneranda et al. (2019) identified chlorite only in LE16-0009 and epidote in LE16-0009 and -0014. The additional detection with MicrOmega FS of chlorite and epidote in sample LE16-0015-C may be due to the detection of isolated grains in this conglomerate, which were not present in the crushed powder due to the sample heterogeneity. Raman spectroscopy detected pyroxene in many samples with (10 out of 17), but FTIR point spectrometry or MicrOmega detected it only on three different samples. Carbonate is only identified in LE16-0014 with MicrOmega, but was also identified in LE16-0009 and -0013 with the FTIR and in LE16-0015 and -0016 with Raman. To be noted, Raman spectroscopy (Veneranda et al., 2019) also identified olivine and oxides like hematite in several samples, that were not noted here, but could have been dissimulated by the spectral slope between 1.0 and 1.7 µm of the spectra showing serpentine (Annex A Figure 5).

| ***Location*** | ***Sample name*** | ***Sample state*** | ***serpentine*** | ***HCP*** | ***chlorite*** | ***epidote*** | ***carbonate*** | ***Additional identified phases in Lantz et al. (2020)*** |
|---|---|---|---|---|---|---|---|---|
| Leka | LE16-0001-B | rock | L | | | | | |
| Leka | LE16-0002-N | rock | L | | | | | |
| Leka | LE16-0003-N | rock | L | | | | | |
| Leka | LE16-0004-D | rock | L | | | | | |
| Leka | LE16-0005-B | rock | L | | | | | |
| Leka | LE16-0006-N | rock | L | | | | | |
| Leka | LE16-0007-B | rock | L | | | | | *HCP* |
| Leka | LE16-0008-B | rock | L | L | | | | |
| Leka | LE16-0009-B | rock | | | L | L | | *carbonate* |
| Leka | LE16-0010-B | rock | L | | | | | |
| Leka | LE16-0011-B | rock | | L | L | | | *hornblende* |
| Leka | LE16-0012-B | rock | L | | | | | |
| Leka | LE16-0013-B | rock | | | | | | *carbonate, hornblende* |
| Leka | LE16-0014-N | rock | | | L | L | L | *hornblende* |
| Leka | LE16-0015-C | rock | L | | | | | |
| Leka | LE16-0016-C | rock | L | | | | | |
| Leka | LE16-0017-B | rock | L | | | | | |

*Table 9. Same as table 3 but for the Leka samples.*

### 4.8. USA-Oregon samples

These samples come from the John Day Formation (Oregon), USA [Retallack et al., 2000]. They were collected at John Day Valley (JD16-001 to -009), Clarno (-016 to -020), and Painted Hills (the others).

The samples from Oregon present a large diversity in hydrated mineral species identified with MicrOmega FS (Table 10). Phyllosilicates were identified in many samples, with spectrally distinguishable smectites, kaolinite (Annex A Figure 7), mica, chlorite, and possible mixed-layer smectite-mica or smectite-chlorite (Annex A Figure 8), using in particular the specific OH combination bands in the 2.2-2.5 µm region (e.g. Bishop et al., 2008). Oxides are also widely present, in particular likely hematite in association with smectite or kaolinite (Annex A Figure 9). Carbonate (Annex A Figure 8) and sulfate (gypsum, Figure 13) were also identified in a few samples, as well as zeolite. Pyroxene was also identified in some samples (Annex A Figure 8), and possibly olivine. JD16-0010-B is shown as an example with detection of kaolinite and hematite in different amount over most of the field of view of MicrOmega FS, and gypsum over a small part of the sample surface (Figure 13).

Mineralogy of the samples from Oregon seems to correlate with the different origins of the samples (show in the “location” column in Table 10). Samples from “Foree” are enriched in zeolite, those from Picture Gorge and Mascal Basin show mainly smectites, while those from Painted Hills and Clarno show the association of hematite with kaolinite or smectite, except for samples JD16-0020-B and -0021-B that show spectra more similar to chlorite or vermiculite.

| ***Location*** | ***Sample name*** | ***Sample state*** | ***Fe/Mg-smectite*** | ***Al-OH smectite*** | ***kaolinite/halloysite*** | ***smectite*** | ***hematite*** | ***undifferentiated Fe oxide*** | ***carbonate*** | ***gypsum*** | ***undifferentiated sulfate*** | ***zeolite*** | ***mica*** | ***vermiculite or mixed-layer smectite-mica*** | ***chlorite*** | ***possible allophane*** | ***LCP*** | ***HCP*** | ***olivine*** | ***Additional phases identified in Lantz et al. (2020)*** |
|---|---|---|---|---|---|---|---|---|---|---|---|---|---|---|---|---|---|---|---|---|
| Foree | JD16-0001-B | rock | | | | | | | | | | L | L | | | | | | | *carbonate* |
| Foree | JD16-0002-B | rock | | | | | | | L | | | L | | | | | | | | *mica* |
| Picture Gorge | JD16-0003-B | rock | | | | | | | | | | | | | | | | | | *Si-OH phase* |
| Picture Gorge | JD16-0004-B | rock | L | | | | | | | | | | | | | | | | | |
| Picture Gorge | JD16-0005-B | rock | | | | | | | | | | | | | | | | L | | |
| Picture Gorge | JD16-0006-B | rock | L | | | | | | | | | | | | | | | | | *Si-OH phase* |
| Picture Gorge | JD16-0007-N | rock | | | | | | | | | | | | | | | | L | ? L | |
| Mascall Basin | JD16-0008-B | rock | | | | L | | | | | | | | | | | | | | |
| Mascall Basin | JD16-0009-B | rock | | | | L | | | | | | | | | | | | | | |
| Painted Hills | JD16-0010-B | rock | | | L | | L | | | | | | | | | | | | | |
| Painted Hills | JD16-0011-N | rock | L | | | L | | | | | | | | | | | | | | *olivine, Si-OH phase* |
| Painted Hills | JD16-0012-B | rock | | | L | | L | | | | | | | | | | | | | |
| Painted Hills | JD16-0013-B | rock | | | L | | L | | | | | | | | | | | | | *carbonate* |
| Painted Hills | JD16-0014-B | rock | | | | L | L | | | | | | | | | | | | | *carbonate* |
| Painted Hills | JD16-0015-B | rock | | | | L | L | | L | | | | | | | | | | | |
| Clarno | JD16-0016-B | rock | | L | | L | L | | | | | | | | | | | | | |
| Clarno | JD16-0017-B | rock | | | | L | L | | | | | | | | | | | | | *kaolinite* |
| Clarno | JD16-0018-B | rock | | | L | L | L | | | | | | | | | | | | | *carbonate* |
| Clarno | JD16-0019-B | rock | | L | | | | | | | | | | | | | | | | *carbonate* |
| Painted Hills | JD16-0020-B | rock | | | | | | | | | | | | | | | | | | *carbonate, Si-OH phase* |
| Painted Hills | JD16-0021-B | rock | | | | | | | | | | | | | L | | | | | |
| Painted Hills | JD16-0022-B | rock | | L | | | L | | | | | | | | | | | | | *kaolinite* |
| Painted Hills | JD16-0023-B | rock | | | L | L | | | | | | | | | | | | | | |
| Painted Hills | JD16-0024-B | rock | | | | L | | | | | | | | | | | | | | *carbonate* |

*Table 10. Same as table 3 but for the USA-Oregon samples.*

When comparing MicrOmega spectral analysis with FTIR point spectrometry analysis (Lantz et al., 2020), we notice full agreement in the detection of high-Ca pyroxene, but low-Ca Pyroxene had not been detected in JD16-0003. There is also agreement in the detection of Fe oxide (probably hematite) except for JD16-0019 and JD16-0023. Concerning hydrated minerals, kaolinite was also identified in the same samples except for JD16-0017 where Lantz et al. (2020) detect a kaolinite signature (possibly thanks to the highest spectral resolution of the FTIR compared with MicrOmega FS, in the presence of smectite/kaolinite mixture with a lower fraction of kaolinite). Smectites and chlorite were identified in the same samples except for additional detections with MicrOmega in JD16-0003, JD16-0007, although traces of hydrated phases were also identified in Lantz et al. (2020) in both samples. Finally, the detection of carbonate with MicrOmega is less obvious, with detection by Lantz et al. (2020) in JD16-0001, -0002, -0013, -0014, -0019, -0020 and -0024, from which JD16-0013, -0018, -0019 and -0024 had carbonate signatures at wavelengths present in the MicrOmega FS range (3.35-3.50 µm), while we detect carbonate with MicrOmega FS only in JD16-0002 and -0015.

Pyroxene detections with MicrOmega FS are confirmed with Raman analysis (Veneranda et al., 2019). They also detected hematite in most samples, but not in JD16-0018 and -0019 where MicrOmega FS identified some and where Raman identified goethite. In the same samples, they also identified serpentine, not detected in this study. These differences could be due to heterogeneity in the sample, the crushed powder analyzed in Veneranda et al. (2019) may have been prepared with a sample part of different mineralogy. Carbonate (calcite) was also identified with Raman but not totally in the same samples (JD16-0001, -0002, -0009, -0020 and -0021). Again, we cannot compare smectite and kaolinite detections as Raman spectroscopy could not identify phyllosilicates in these samples.

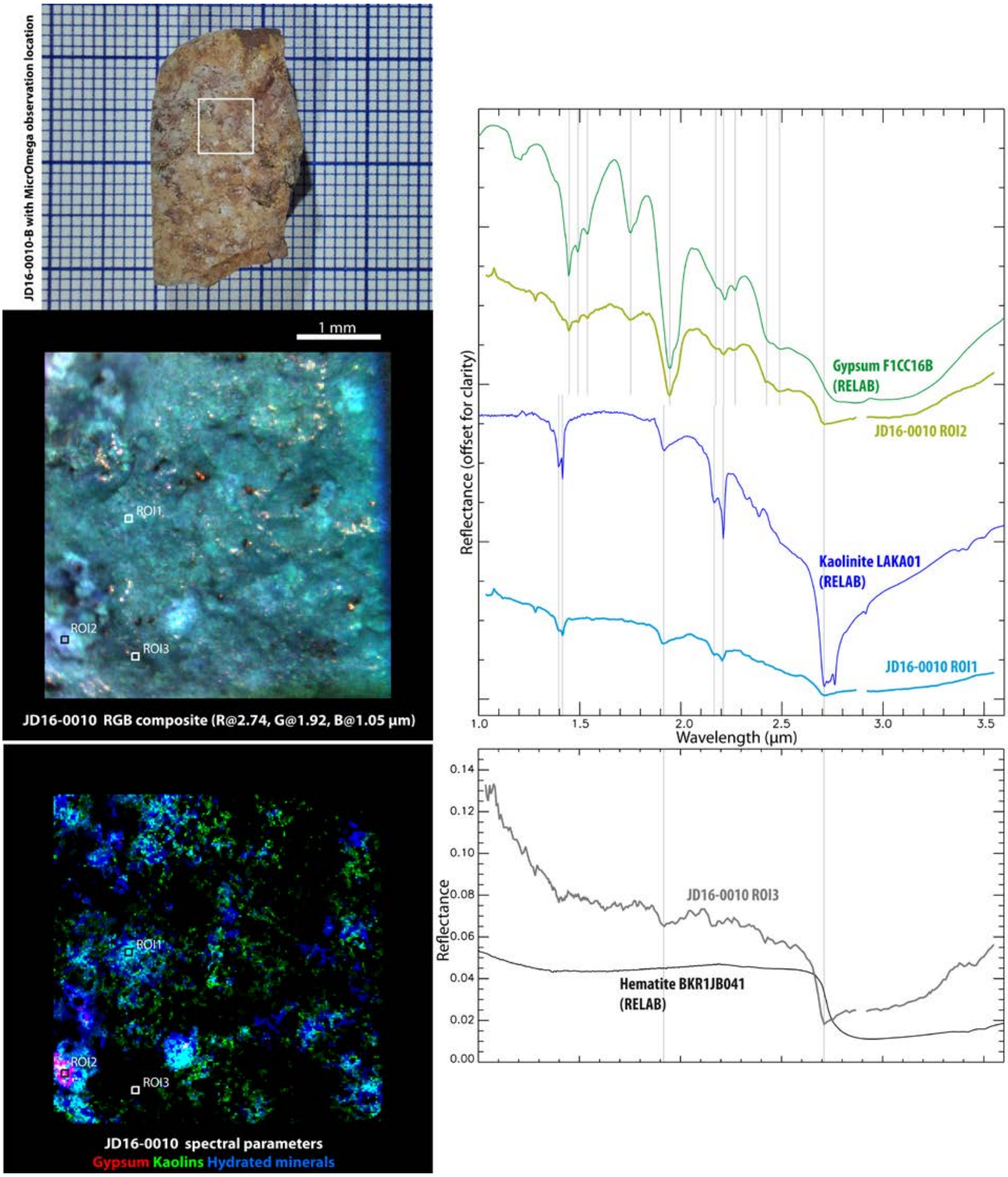

*Figure 13. Same as Figure 6 but for sample JD16-0010-B.*

### 4.9. Canary Islands samples

The samples were picked at various location in the Grand Canary and Tenerife islands [Troll and Carracedo, 2016; see Dypvik et al., 2021, for a more detailed description of the sampling sites]. Similar first two letters show a common location of sampling, for example MR16-0001 and MR16-0002 were both collected in “Mna Reventada” outcrop.

Those samples show a large geological diversity, with igneous rocks of different compositions at different stages of alteration, as well as sandstones. Table 11 displays all detections and tentative detections of Tenerife and Grand Canary samples, and reveals the high mineral diversity of these samples.

Sandstones AM16-0001-P and -0002-B display a variety of minerals and both have sulfate or halite (gypsum for AM16-0002-B and either a polyhydrated sulfate or halite for AM16-0001-P). Detections of hydrated minerals in sample AM16-0002-B are illustrated in Figure 14.

Olivine was detected in various samples (Annex A Figure 10). Basaltic glass is possibly detected in volcanic rocks CB16-0001 and AG16-0001 (Annex A Figure 11). Iron oxides and oxide-hydroxides are also likely detected in various samples (Annex A Figure 12). Finally carbonate (Annex A Figure 13) was also identified in RN16-0001-B (possible surface deposit) and in the volcanic phonolite rock AD16-0001-B (in grains embedded in the rock).

| ***Location*** | ***Sample name*** | ***sample state*** | ***kaolinite/halloysite*** | ***Al-OH smectite*** | ***Fe/Mg-OH smectite*** | ***smectite*** | ***smectite and/or zeolite*** | ***zeolite*** | ***vermiculite or mixed-layer smectite-mica*** | ***hydrated silica*** | ***Fe oxide hydroxide*** | ***hematite*** | ***carbonate*** | ***olivine*** | ***pyroxene*** | ***basaltic glass*** | ***gypsum*** | ***sulfate or halite*** | ***mica*** | ***unidentified hydrated species*** | ***additional phases identified in Lantz et al. (2020)*** |
|---|---|---|---|---|---|---|---|---|---|---|---|---|---|---|---|---|---|---|---|---|---|
| Tenerife | MR16-0001-B | rock | | | | | | | | | | | | | | | | | | | |
| Tenerife | MR16-0002-B | rock | | | | | | | | | | | | | | | | | | | |
| Tenerife | AD16-0001-B | rock | | | | | | | | | | | L | | | | | | | | |
| Tenerife | AM16-0001-P | powder | | | | | | | | | | | | | | | | L | | | |
| Tenerife | AM16-0002-B | rock | | | | | L | | | | ? | | | | | | | | | | |
| Tenerife | TF16-0002-C | rock | | L | | | | | | | | | | | | | | | | | |
| Tenerife | TF16-0028-B | rock | | | | | | | | | | | | | | | | | | | *Fe oxide* |
| Tenerife | TF16-0059-C | rock | | | | | | | | | | L | | | | | | | | | *Fe oxide* |
| Tenerife | TF16-0066-B | rock | | | | | | | | | | ? | | | | | | | | | |
| Grand Canary | AG16-0001-B | rock | | | | | | | | | | | | L | L | | | | | | |
| Grand Canary | TO16-0001-B | rock | | | | | | | | | | | | | | ? | | | | | |
| Grand Canary | BT16-0001-B | rock | | | | | | L | | | | | | | | | | | | | *Fe/Mg-smectite, Si-OH phase* |
| Grand Canary | BT16-0002-B | rock | | | L | | | | | | | | | | | | | | | | *Si-OH phase* |
| Grand Canary | CB16-0001-N | rock | | | | | | | | | | ? | | ? L | | ? | | | | | *HCP* |
| Grand Canary | FA16-0001-B | rock | | | | | | | L | | | | | | | ? | | | | | *chlorite, zeolite, carbonate* |
| Grand Canary | FA16-0002-B | rock | | | | | | | | | | | | | | | | | | | *chlorite, smectite, zeolite* |
| Grand Canary | FA16-0003-B | rock | L | | | | | | | | | | | | | | | | ? | | *chlorite, zeolite, carbonate* |
| Grand Canary | RN16-0001-B | rock | | | | | | | | | | | | | L | | | | | | |

*Table 11. Same as table 3 but for the Canary Islands samples.*

The higher mineral diversity compared with other sampling regions makes comparison with other techniques more complex, with a comparison sample by sample. Concerning samples from Tenerife: The MR16 samples showed no spectral signature with the FTIR (Lantz et al., 2020), MicrOmega FS identified a possible Fe oxide hydroxide, while Raman (Veneranda et al., 2019) identified an Fe-oxide (hematite). In sample AD16-0001, carbonate is identified with the three techniques. For the AM16 sandstones, the possible sulfate identified in AM16-0001 with MicrOmega may be the gypsum detected with the FTIR, and the kaolinite in AM16-0002 may be the Al-OH signature seen with the FTIR. In the TF16 samples, FTIR detected hydrated signatures in all samples that may correspond to the Al-OH and other smectites detected here, while the hematite detected in TF16-0002, -0059 and -0066 was also identified with Raman spectroscopy.

Concerning samples from Grand Canary: in AG16-0001, the three techniques identified olivine and pyroxene. In TO16-0001, the signature from a hydrated phase detected with the FTIR (Lantz et al., 2020) may correspond to the hydrated silica identified here. In the BT16 samples, FTIR also identified a zeolite in BT16-0001 and Fe/Mg-smectites in BT16-0002, and Raman (Veneranda et al., 2019) also identified olivine and pyroxene in BT16-0002. In CB16-0001, FTIR and Raman also both detected olivine, and Raman also detected hematite, but no glass. In FA16-0001, FTIR also detects possible chlorite, and in FA16-0003, kaolinite and carbonate. The presence of mica in any FA16 samples has been reported in Lantz et al. (2020) or Veneranda et al. (2019). Finally, in RN16-0001, pyroxene has been detected by all three techniques, olivine has also been identified with Raman, and carbonate is only detected with MicrOmega FS, but on a small spot of the sample surface (<1%).

72

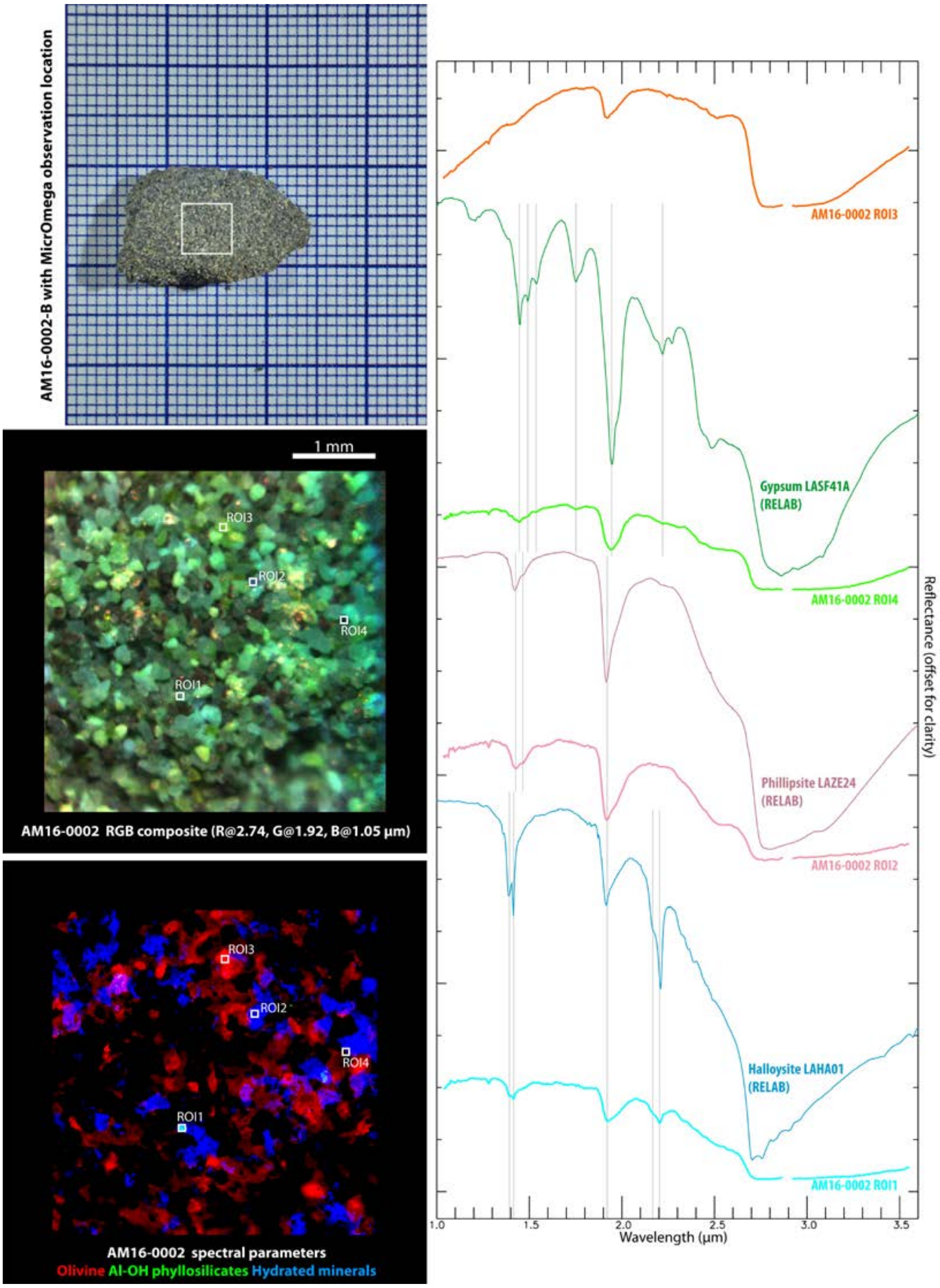

73 *Figure 14. Same a Figure 6 but for sample AM16-0002-B.*

74

75

## 4.10. Otago Samples

Those rocks were sampled in the Blue Spur conglomerate, deposits formed by alteration of the Otago schists geological formation, South Canterbury, New Zealand. Rocks of this location contain vermiculite which has been characterized as trioctahedral, and rich in to $Fe^{2+}$ (Craw et al. 1984). This characteristic makes it potentially correspond to the vermiculite that has been reported in the Oxia Planum region, the future landing site of the ExoMars 2022 mission (Quantin et al., 2020; Krzesińska et al., submitted).

Those samples show evidence of intense alteration, with detection of phyllosilicates, zeolite and potentially iron oxide (Figure 15, Figure 16, Table 12). The presence of a strong spectral slope up to 1.8 µm, a strong 1.9 µm band, a strong drop after 2.2 µm with two bands at 2.25 and 2.33 µm could indicate the presence of mixed-layer chlorite-smectite, rich in Fe/Mg (Figure 15). Study of these samples presented by Krzesinska et al. (submitted), shows that they contain chlorite and vermiculite, that with progress of alteration alters to interstratified illite-vermiculite and Fe-oxide.

Raman and NIR characterization of those samples were not included in the previous studies from Veneranda et al. (2019) and Lantz et al. (2020) as those samples were added lately in the PTAL rock library.

| ***Location*** | ***Sample name*** | ***sample state*** | ***illite*** | ***chlorite*** | ***mixed layered smectite-chlorite*** | ***zeolite*** | ***iron oxide*** |
|---|---|---|---|---|---|---|---|
| Otago | OT-0001-B | rock | | | | | |
| Otago | OT-0002-B | rock | | | | | |
| Otago | OT-0003-B | rock | | | | | |
| Otago | OT-0004-B | rock | | | | | ? |
| Otago | OT-0005-B | rock | | | | | |

*Table 12. Same as table 3 but for the Otago samples.*

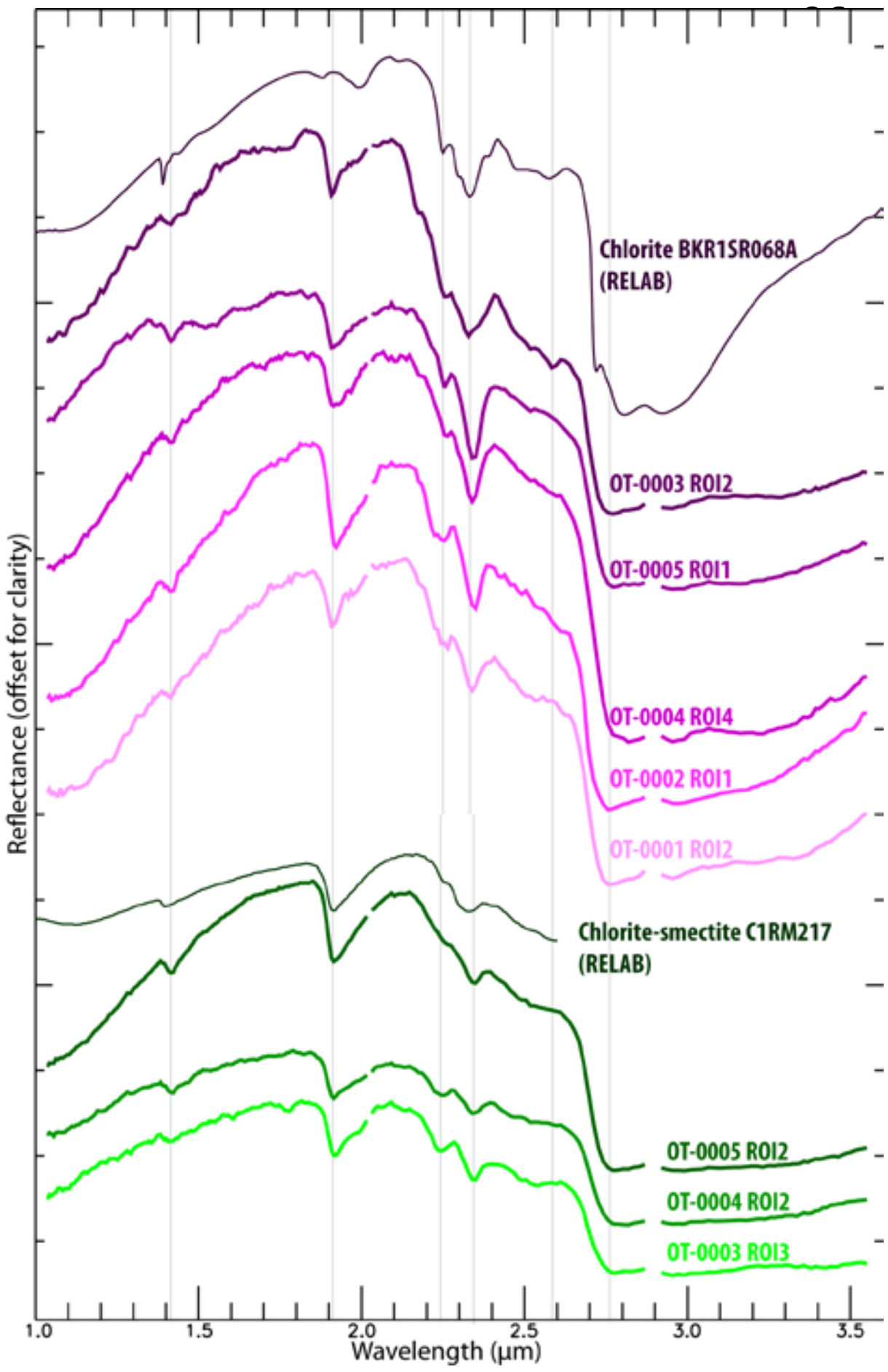

*Figure 15. Selection of local averages of MicrOmega spectra from Otago rock samples compared with reference spectra of chlorite and chlorite-smectite.*

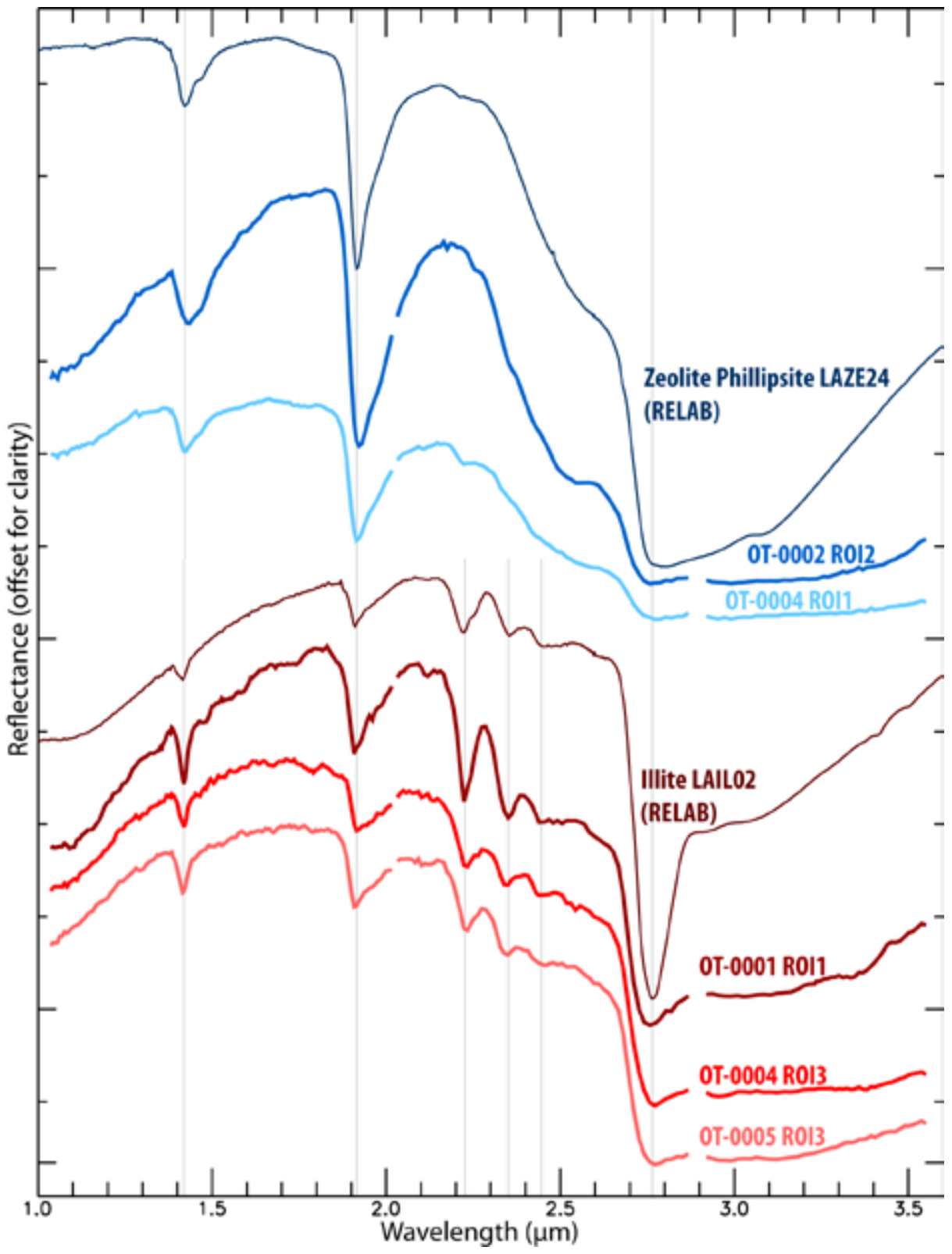

*Figure 16. Selection of local averages of MicrOmega spectra from Otago rock samples compared with reference spectra of illite and zeolite.*

## 4.11. Lonar Crater Samples

| Location | Sample name | sample state | HCP | olivine | Fe/Mg-phyllosilicates | glass | iron oxide | organics | unidentified hydrated species |
|---|---|---|---|---|---|---|---|---|---|
| Lonar | LO-0001-B | rock | | | | ? | | | |
| Lonar | LO-0002-B | rock | | | | | | | |
| Lonar | LO-0003-B | rock | | | | ? | | | |

*Table 13. Same as table 3 but for the Lonar crater samples.*

Those samples were collected at the Lonar impact crater site, Buldhana district, Maharashtra, India. Lonar crater is a simple crater formed in the basaltic target of the Deccan large igneous province (Fredriksson et al., 1973; Senthil Kumar et al., 2014). Being one of only three preserved terrestrial craters formed in basaltic target, Lonar is an important analogue for studies of shock processes in the inner solar system and for post-impact alteration triggered in basaltic rocks. Raman and NIR characterization of those samples were not included in the previous studies from Veneranda et al. (2019) and Lantz et al. (2020) as those samples were added lately to the PTAL rock library.

Samples LO-0001 and -0003 exhibit partial alteration with the presence of pyroxene, olivine in one of the samples, together with Fe/Mg-phyllosilicates. Basaltic glass in suspected also in those samples. LO-0002 shows the presence of pyroxene and iron oxide together with organics (Figure 17).

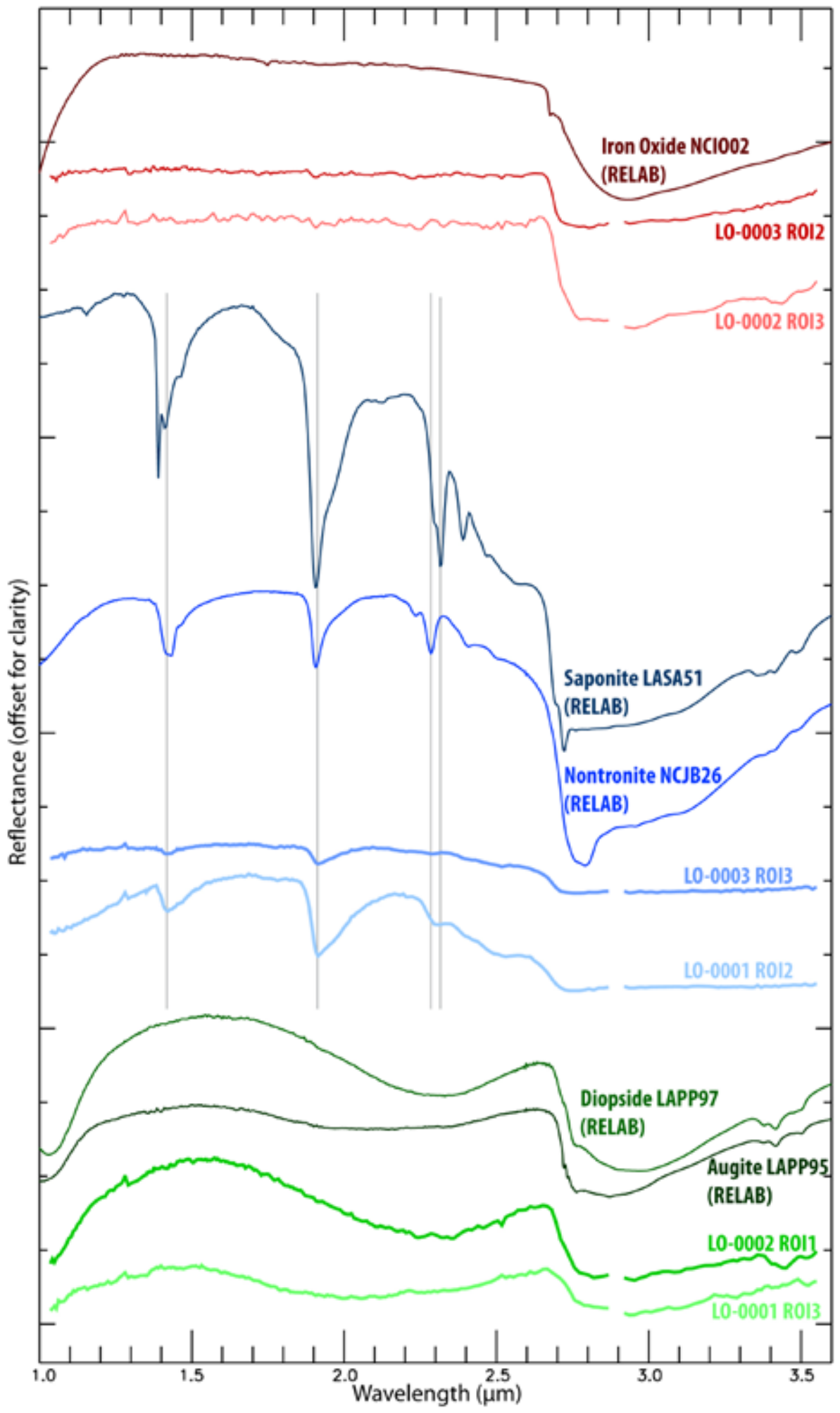

*Figure 17. Selection of local averages of MicrOmega spectra from Lonar crater rock samples compared with reference spectra. The small spike in some spectra around 1.3 µm is an artefact.*

## 5. General observations

### 5.1. Sample mineralogy

Minerals and mineral species detected with MicrOmega in the PTAL samples include: Olivine, High Calcium Pyroxene (as Diopside and Augite), Low Calcium Pyroxene (as Pigeonite, Enstatite or Hyperstene), Amphiboles (Actinolite, Hornblende), Epidotes (Epidote, Zoisite), Zeolites, Opals, Phyllosilicates (Serpentine, Chlorite, Kaolinite, Smectites, Illite, Mica…), Oxides and Hydroxides (Hematite, Goethite and Lepidocrocite, Diaspore), Carbonates, and Sulfates (Gypsum, Jarosite, Copiapite…).

A first look at the mineralogical characterization with MicrOmega FS reveals that their mineralogy is in general agreement with the characteristic of the samples. Samples from Oregon-USA show a strong signature suggestive of abundant hydrated silicates and oxides, which is expected for those intensively weathered volcanic rocks. On the other hand, samples from Iceland show mostly olivine, pyroxene and glass, in agreement with the nature of those fresh volcanic rocks. Samples from Leka in Norway show presence of serpentine and/or chlorite, expected for those deeply altered ophiolite complex (Dypvik et al., 2021).

Spectral parameter maps enable a rough estimate of the concentration of the detected minerals at the surface of the different samples (as illustrated in Figures 4, 6, 10, 12, 13 and 14) and show the spatial relation between minerals. For example an impact breccia show well spatially delimited grains of very different mineralogy (Figure 6), while deeply altered rocks show more diffuse and intricate relation between the crystals (Figure 12).

### 5.2. Organic matter detection

The PTAL samples analyzed with MicrOmega are all from natural rocks, so contamination with organic matter is expected. Figure 18 shows some examples of potential detection of organic molecules on the surface of samples in the spectral domain 3.1-3.55 µm. C-H stretching modes in aliphatic molecules occurs in this spectral domain (e.g. Clark et al., 2009). Spectra likely indicate the presence of different molecules with aliphatic C-H. The organic matter is probably a mixture of different aliphatic molecules as the spectral bands are less sharp than for the reference spectra where only one type of molecule is present in the rock.

In some cases, MicrOmega observations have been made on rock surfaces that have been exposed to the surface for years. For example, sample RU16-0001-B (see figure 4) presents spectra that fit reference spectra of dry vegetation (Figure 19). The presence of an ancient lichen is a possibility.

Concerning the spatial presence of organic molecules within the samples, for most samples where spectral bands were detected in the 3.3-3.5 µm, they were not detected in all regions of interest at the surface of the samples, but only in few ones. The only exception is sample LO-0002-B from Lonar crater were a large 3.45 µm band is present on most of the surface (Figure 18). In addition, while Raman spectroscopy did detect carbon in some of the PTAL samples, it did not detect it in the samples where we suspect the presence of organic compounds with MicrOmega (Veneranda et al., 2019). The presence of rare and small areas enriched in organic compounds may explain this difference.

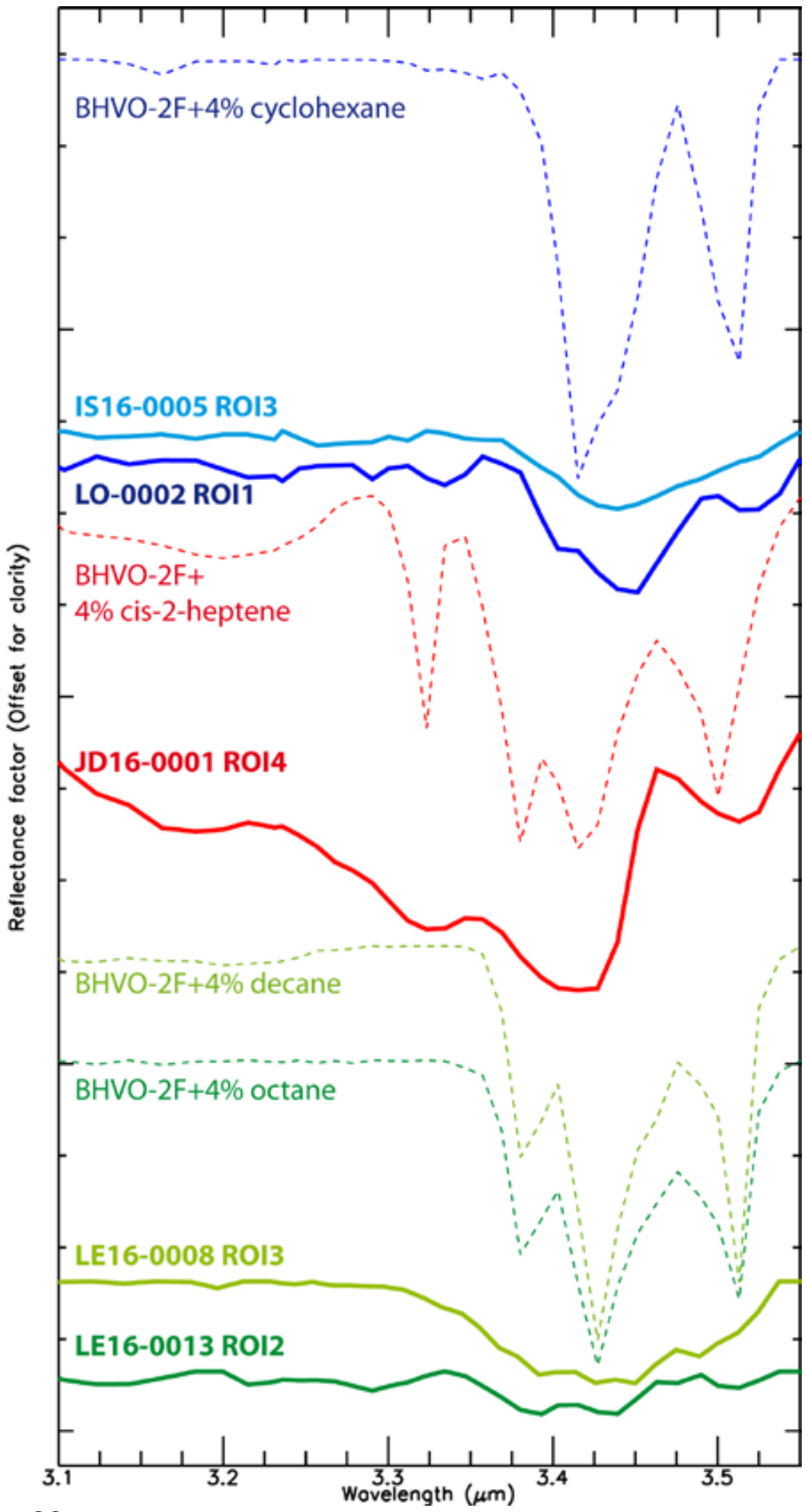

*Figure 18. Examples of potential detections of organic matter in PTAL samples, and comparison with reference spectra of mixtures of basaltic rocks with organic molecules. All spectra are continuum-removed. The reference spectra in dashed lines are extracted from the USGS spectral library (Kokaly et al., 2017; BHVO-2F is a USGS Basalt standard) and have been resampled to the MicrOmega spectral sampling. Reference spectra illustrates the type of molecules or mixture of molecules that may explain the MicrOmega spectra but are not definitive identification of these particular molecules.*

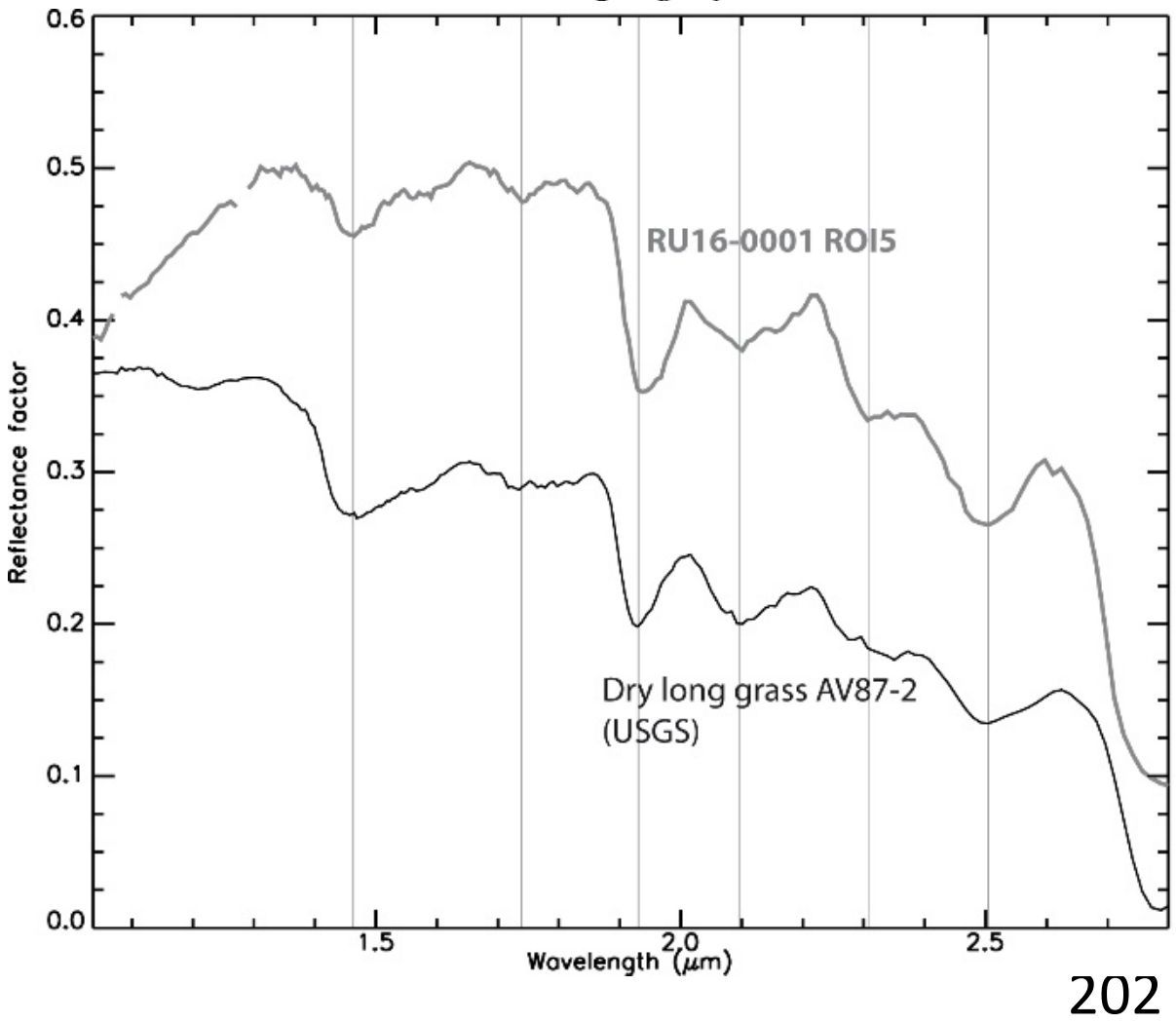

*Figure 19. Example of detection of dry vegetation at the surface of a sample after years of surface exposition.*

## 5.3. General comparison between MicrOmega FS and an FTIR point spectrometer

Figures 6, 7 and 9 allow a visual comparison between spectra from MicrOmega FS and spectra from the point spectrometer used in Lantz et al. (2020) on the same crushed powder samples. The spectral resolution is different (20 $cm^{-1}$ for MicrOmega FS and 4 $cm^{-1}$ for the point spectrometer) but the resolution of the figures in not sufficient to make this difference visible. Also, please note that the geometry of illumination/observation is different between MicrOmega FS and the point spectrometer, leading to potential differences in spectral continuum level and depth of spectral bands.

Besides, although the same crushed powders were used for both instrument, those powders have been put away in vials between the two analyses, so the characterizations were not made on the exact same grains: some minor species could be absent in one or the other of the analyses.

The first observation is that the average MicrOmega FS spectrum over most of the FoV (2.6 mm x 2.6 mm) and the point spectrometer spectrum (FoV diameter < 1mm) are in good agreement in both cases, except for the 2.72 µm band of sample BR16-0001 (Figure 9) which is significantly deeper with the point spectrometer. However, we can notice that the point spectrometer spectra of those two examples do not allow the identification of all the different mineral species that we could identify in this study with MicrOmega FS on the same crushed powders. Pyroxene and zeolite could not be identified in VO16-0002 (Figure 7), and chlorite and Fe/Mg-smectites could not be identified in BR16-0001 (Figure 9) with the spectra from the point spectrometer.

Tables presenting MicrOmega FS mineral identification for all samples enable to compare results between MicrOmega FS and the point spectrometer. Here again, we identify three reasons for observed differences in mineral identification:

- A recurring difference comes from the more extended spectral domain used with the point spectrometer (0.8-4.2 µm), which leads to additional detections that cannot be made with MicrOmega FS (~0.99-3.6 µm). In particular, detection of carbonates with the spectral band around 4.0 µm, which has often been made only with this band in Lantz et al. (2020), is impossible with MicrOmega FS. This suggest that the carbonate detection is disturbed by the water absorption band 3.0-3.6 µm in our study of natural Earth samples. With dryer samples, as samples on Mars are expected to be, the identification of the 3.4 µm band of carbonates may be easier. In addition, the identification of Fe oxides around 1.0 µm is much more difficult with spectra that starts at 1.0 µm.

- The spectral resolution and signal-to-noise ratio of the point spectrometer enables in some cases to identify minerals with thin and shallow spectral bands that could not be detected with MicrOmega FS.
- Another source of differences comes from the heterogeneities of the samples, as illustrated in Figures 6 and 7. In general, there is a higher chance of identifying a minor species in some of the pixels of the 25 mm² FoV of MicrOmega FS than in the single ~1 mm² FoV of the point spectrometer.

### 5.4. Bulk rock samples vs. crushed powder

The last section discusses comparison of analyses of the same crushed powders, but in this study, we chose to characterize mainly exposed surfaces of rock sub-samples. Figures 6, 7 and 9 compare the observations made on those two types of sample preparation.

Comparison between Figures 6 and 7 shows that the analysis of bulk rock surface of VO16-0002 enables the identification of more species than the crushed powder that shows only very few large grains (>100 µm): carbonate and the variety in hydrated minerals cannot be identified in the powder. Figure 9 shows that in the case of a coarser powder (several grains >100 µm) for sample BR16-0001, the same minerals can be identified, although the spectral signatures are weaker with the crushed powder than with the bulk rock surfaces in the case of chlorite or actinolite. Those examples reveal a more general observation made from this study, that fine crushed powders (very few grains >100 µm) reveal less mineral diversity than coarser powders and bulk rock surfaces. For comparison, the crushing station onboard the Rosalind Franklin rover of the ExoMars mission will deliver powders with a significant portion of grains several 100’s µm in size (Redlich et al., 2018).

It is also worth highlighting that MicrOmega being an imager, and the analysis being performed without destruction of the rock’s natural structure, the original distribution of the minerals in the rock and their textural relationship one to another could be imaged, and a wider spectral diversity could be identified.

### 5.5. Products available in the online PTAL database

Products linked to the PTAL sample characterization with MicrOmega FS will be included in the PTAL database (database link available on the PTAL project website www.PTAL.eu).

Being a hyperspectral imager, MicrOmega data are heavy and complex to present. The whole spectral cubes are not for download on the online database, but we extracted images, maps and spectra to support the mineral interpretation that is made and also detailed on the online database. For each MicrOmega observation, are made available:

- a picture of the sample with exact location of the MicrOmega FS observation,
- one RGB image of the observation extracted from LED-illuminated images,
- one RGB image of the observation extracted from AOTF-illuminated images,
- the same image as above with location of ROIs for the available spectra,
- a large spectral average of the FoV of the observation,
- a few selected local spectral averages of identified spectral endmembers in ROIs of the observation,
- a few spectral parameter maps that were used to make the analysis of the observation.

All products are available for online display, can be zoomed-in and are downloadable.

## Acknowledgments

We are grateful for the whole team that enabled the development of the MicrOmega instrument and its laboratory facilities. We also wish to thank the people who sampled the PTAL rocks that were characterized here: H. Hellevang, C. Sætre, F. Torrado, V. Troll, A. Caramazana, J.-C. Viennet, D. Jerram, A. Crosta, D. Craw and D. Ray. This research utilizes spectra acquired by E.A. Cloutis, J.L. Bishop, C.M. Pieters, J.R. Michalski, J. Mustard and R.V. Morris with the NASA RELAB facility at Brown University, as well as spectra from the USGS spectral library splib07a (Kokaly et al., 2017). The authors also wish to thank T. Glotch and J. Michalski for their careful corrections and help to improve the paper.

The PTAL project is funded through the European Union's Horizon 2020 Research and Innovative Program H2020-COMPET-2015 (grant 687302).

*ANNEX to "*PLANETARY TERRESTRIAL ANALOGUES LIBRARY PROJECT: 3. CHARACTERIZATION OF SAMPLES WITH MICROMEGA", Loizeau et al.

## Section 4.5. Oslo rift samples, supplementary figures

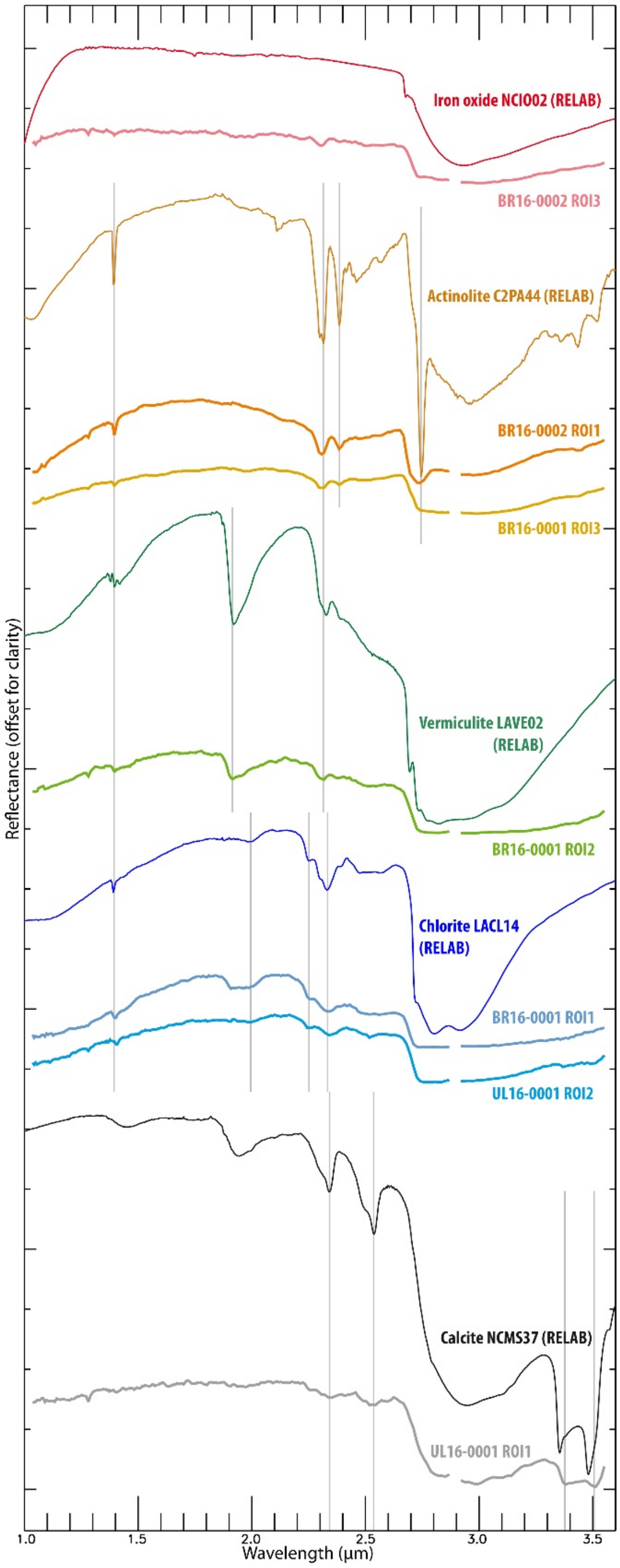

*Figure 1. Selection of local averages of MicrOmega spectra from Oslo rift rock samples compared with reference spectra. The small shift between the calcite spectral bands around 3.4 and 3.5 µm and the MicrOmega FS spectrum from UL16-0001 ROI1 might be due to a slightly different composition than calcite, but the spectral calibration of MicrOmega FS is less constrained in this spectral region, which could also explain this shift.*

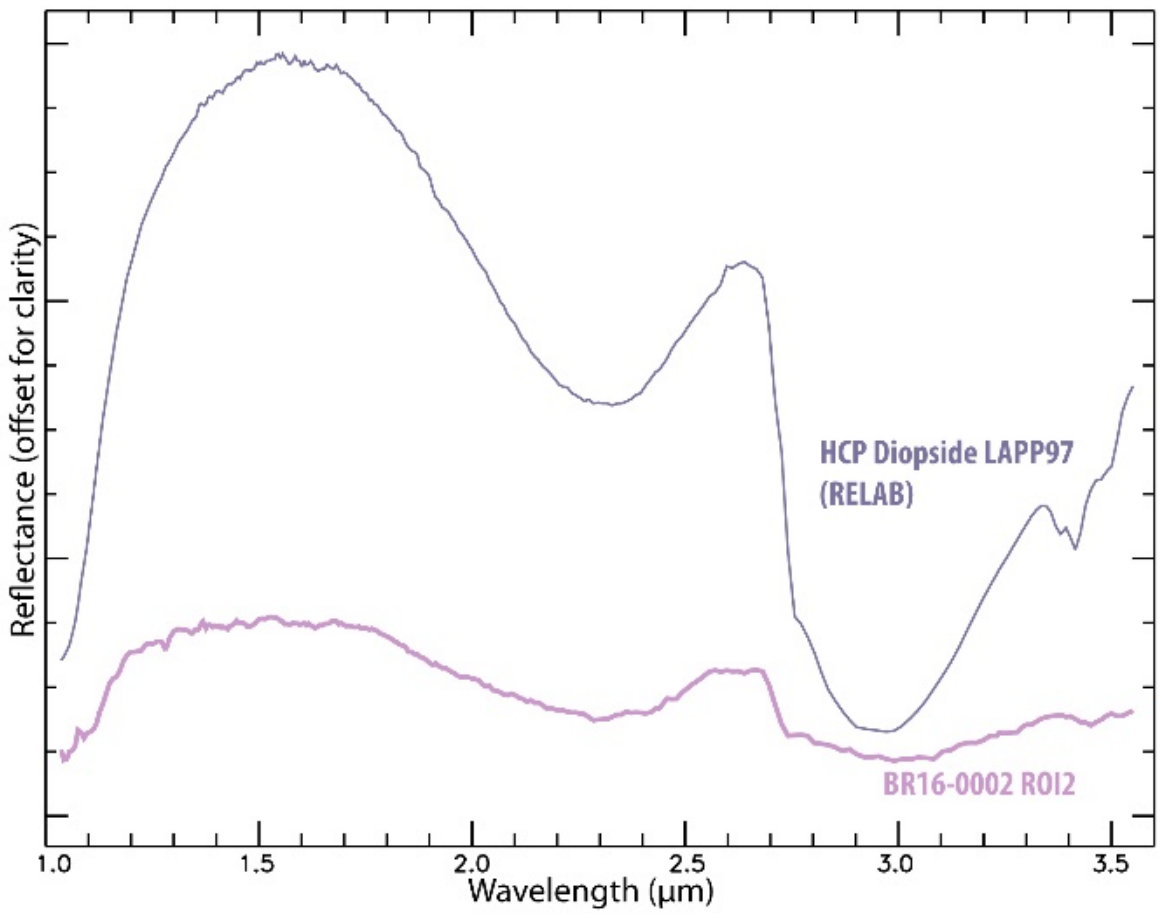

*Figure 2. Local average of MicrOmega spectra from Oslo rift rock sample BR16-0002 compared with reference spectrum of High-Calcium-Pyroxene Diopside.*

## Section 4.6. Iceland samples, supplementary figures

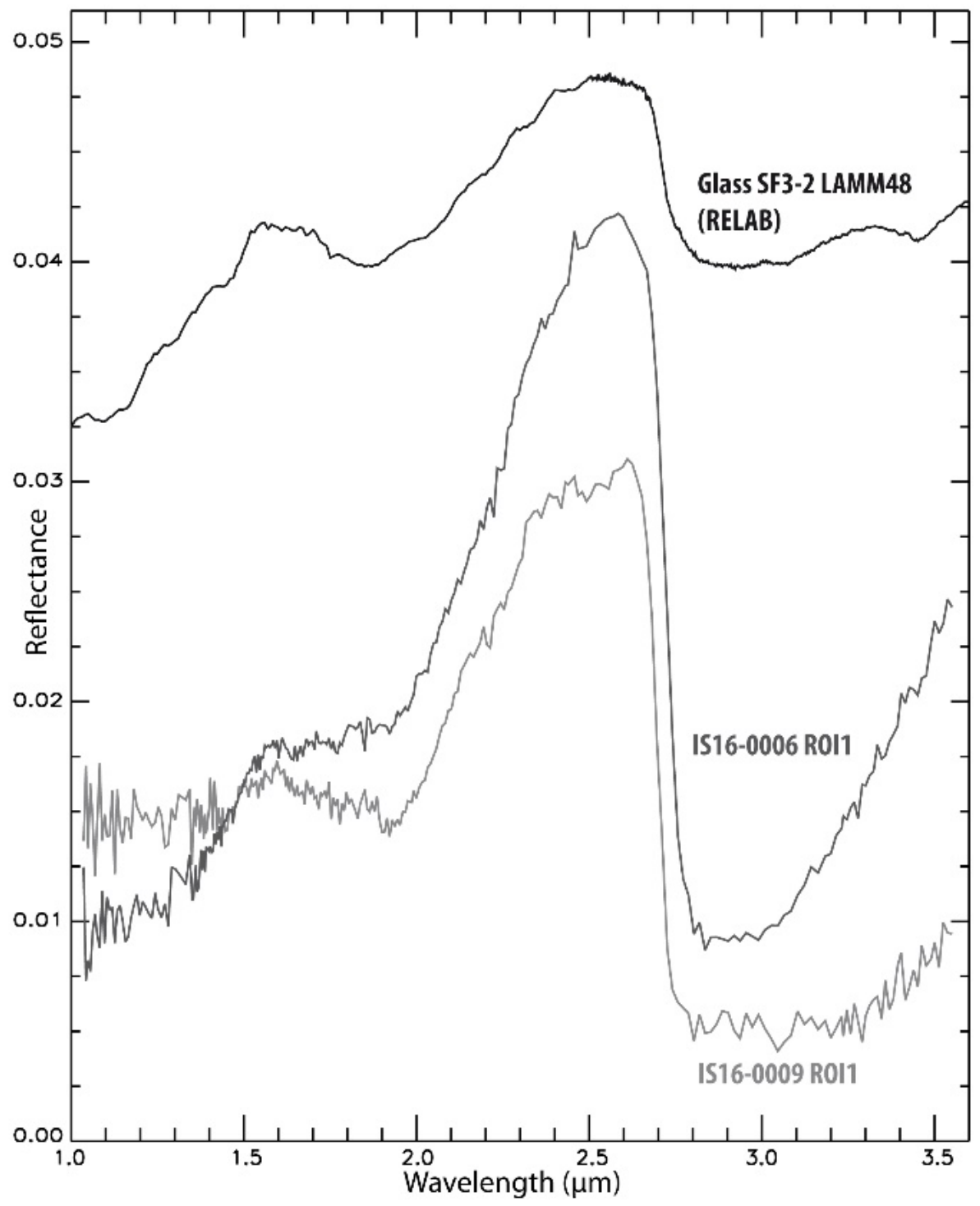

*Figure 3. Selection of local averages of MicrOmega spectra from Iceland rock samples compared with reference spectra of basaltic glass.*

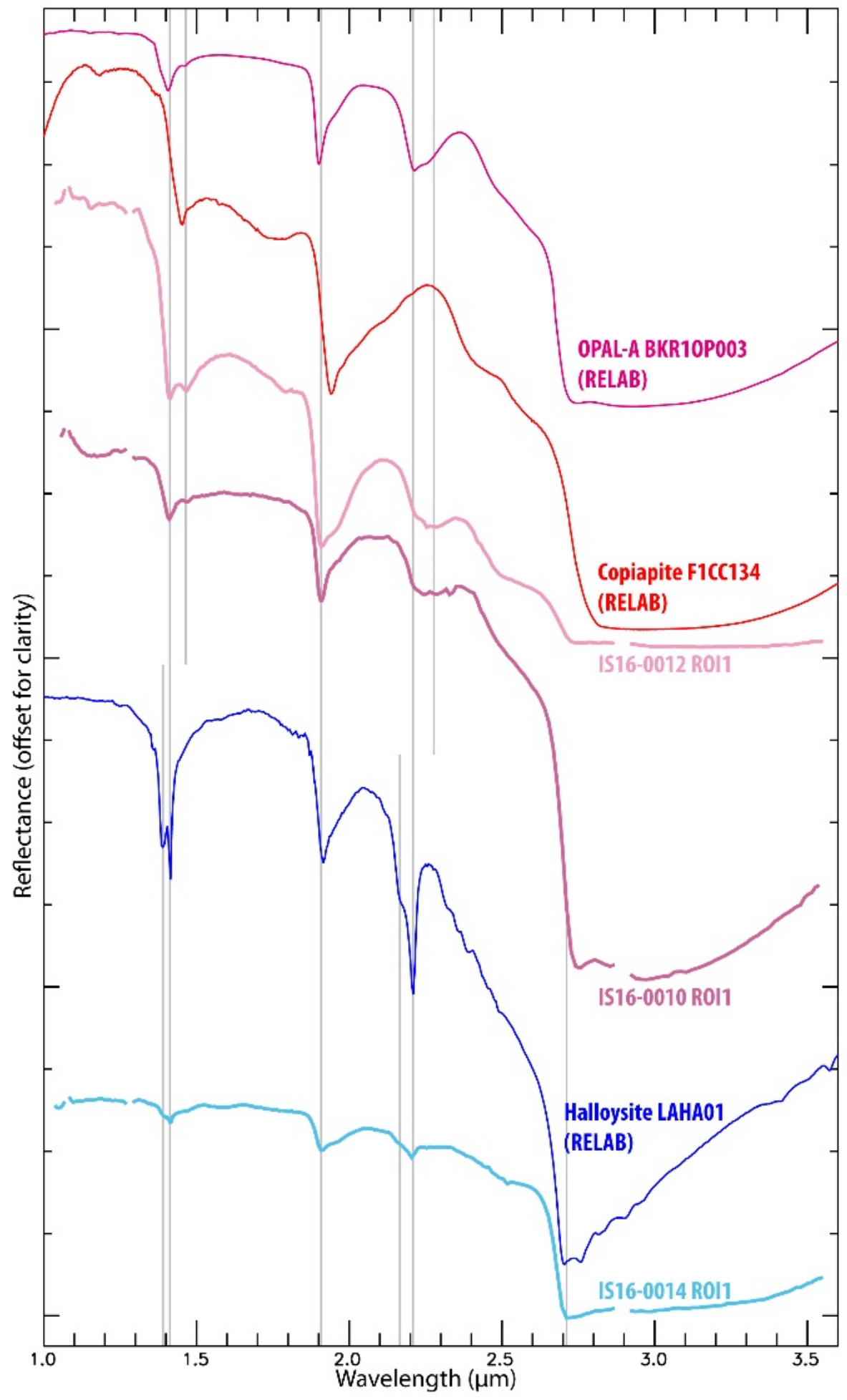

*Figure 4. Selection of local averages of MicrOmega spectra from Iceland solfatara rock samples compared with reference spectra*.

## Section 4.7. Leka samples, supplementary figures

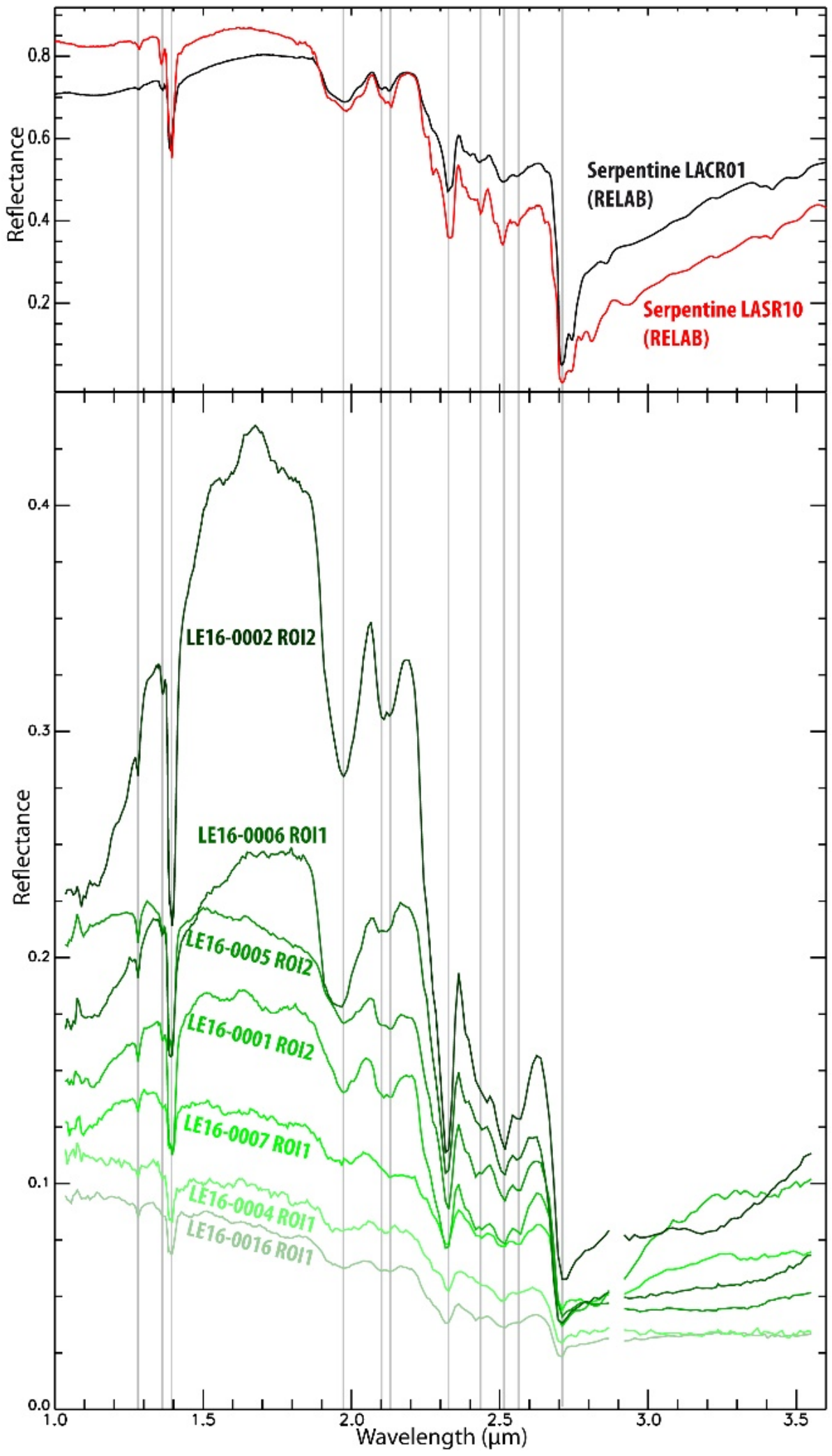

*Figure 5. Selection of local averages of MicrOmega spectra from Leka rock samples compared with reference spectra of serpentine.*

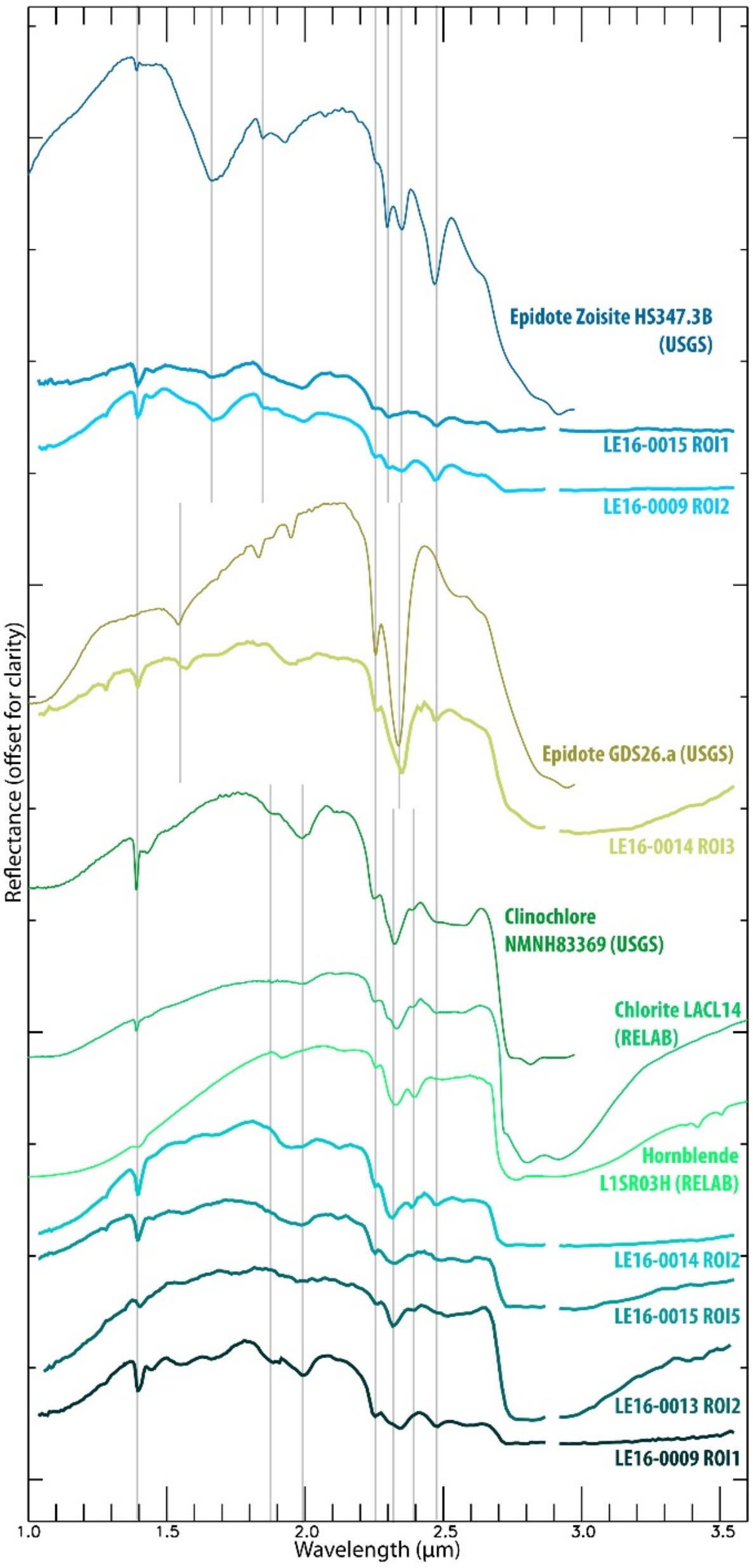

*Figure 6. Selection of local averages of MicrOmega spectra from Leka rock samples compared with reference spectra.*

## Section 4.8. USA-Oregon samples, supplementary figures

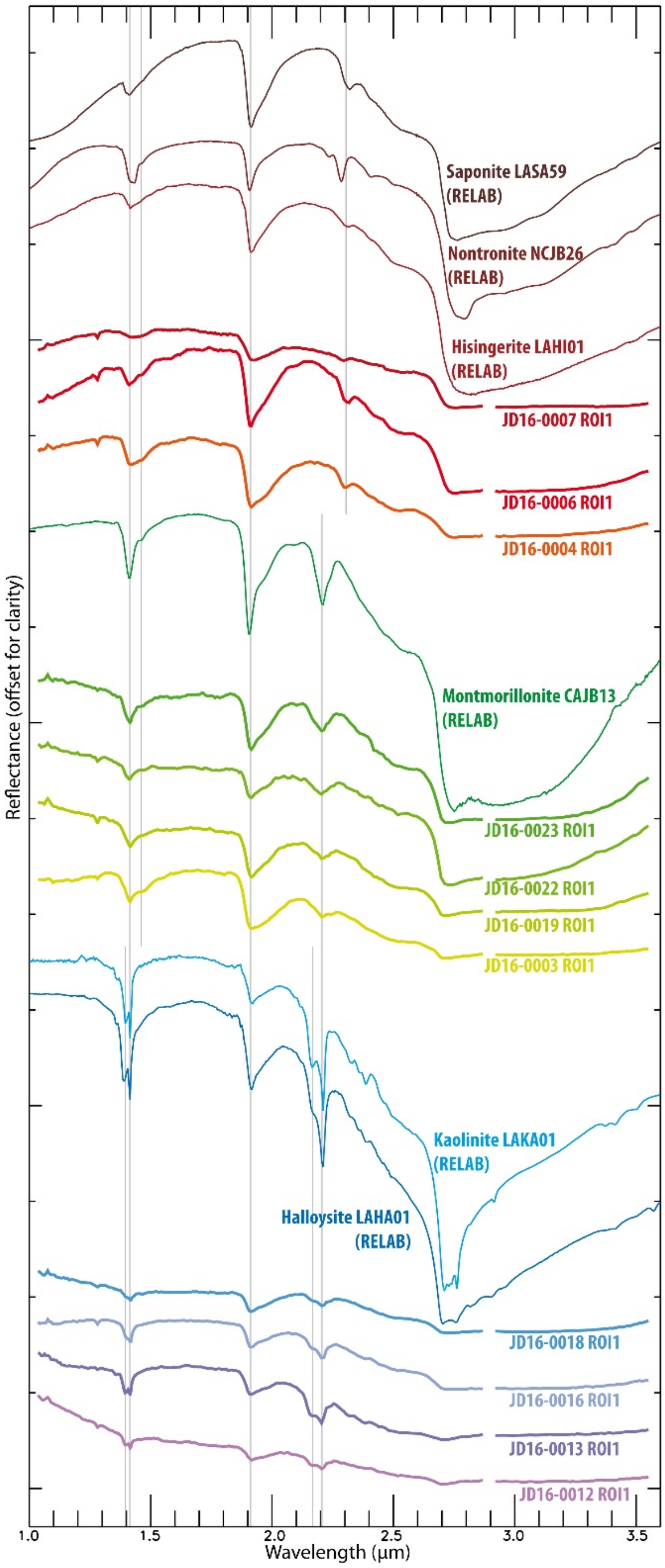

*Figure 7. Selection of local averages of MicrOmega spectra from Oregon rock samples compared with reference spectra of phyllosilicates.*

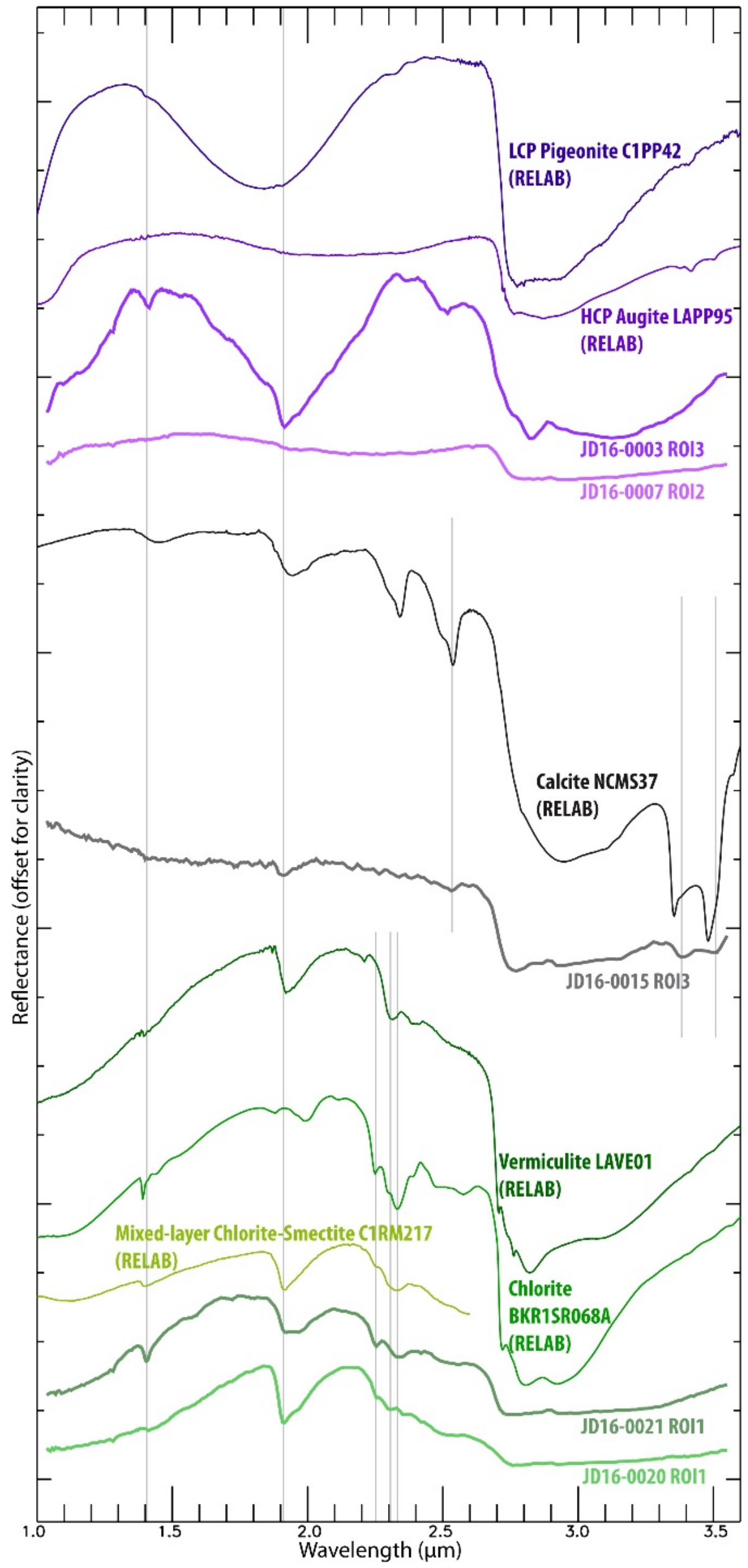

*Figure 8. Selection of local averages of MicrOmega spectra from Oregon rock samples compared with reference spectra. The small shift between the calcite spectral bands around 3.4 and 3.5 µm and the MicrOmega FS spectrum from JD16-0015 ROI3 might be due to a slightly different composition than calcite. This shift could also be the result of the spectral calibration uncertainty of MicrOmega in this wavelength range.*

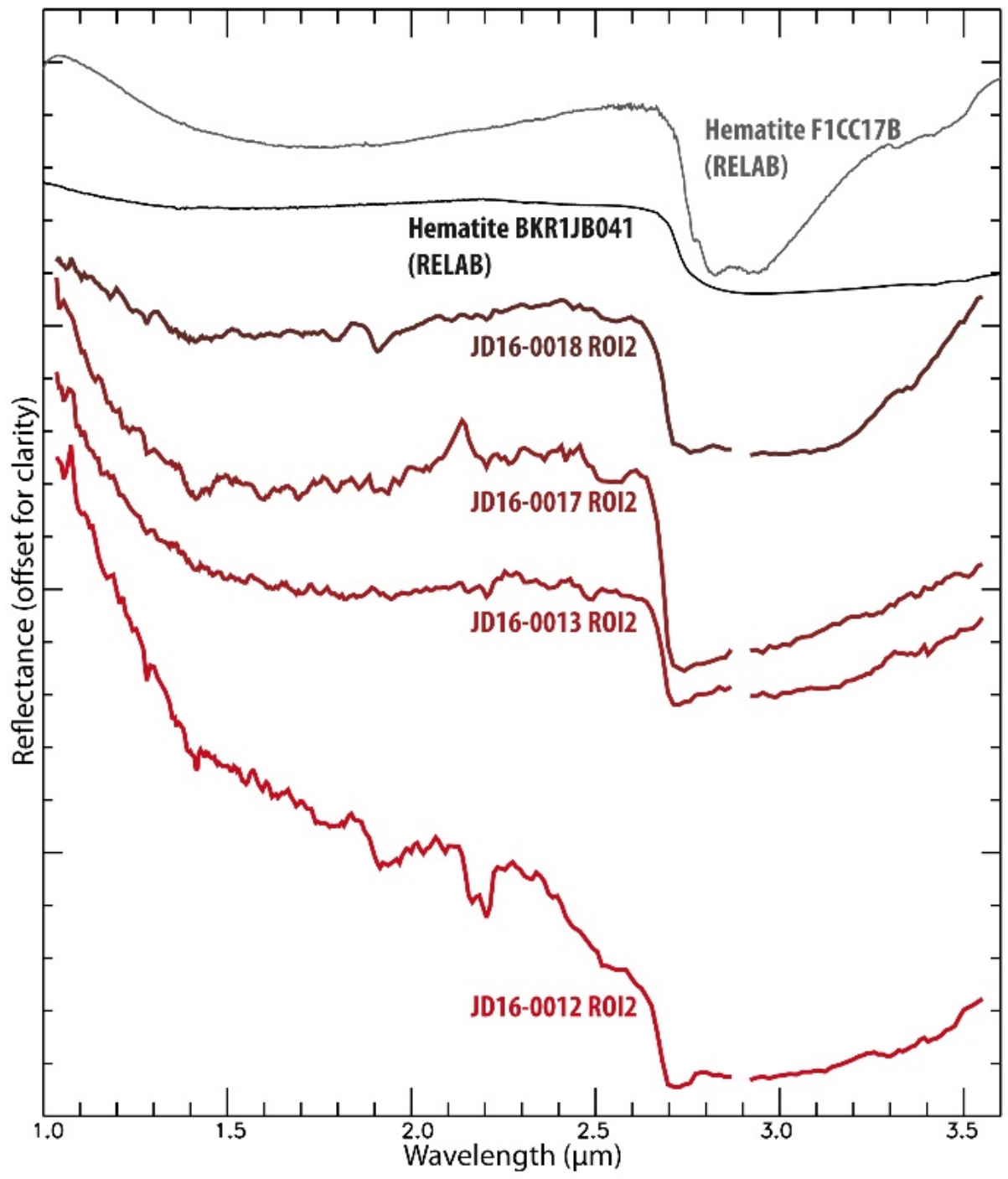

*Figure 9. Selection of local averages of MicrOmega spectra from Oregon rock samples compared with reference spectra of hematite. High noise in these spectra is due to the low albedo of these grains (~10% and below).*

## Section 4.9. Canary Islands samples, supplementary figures

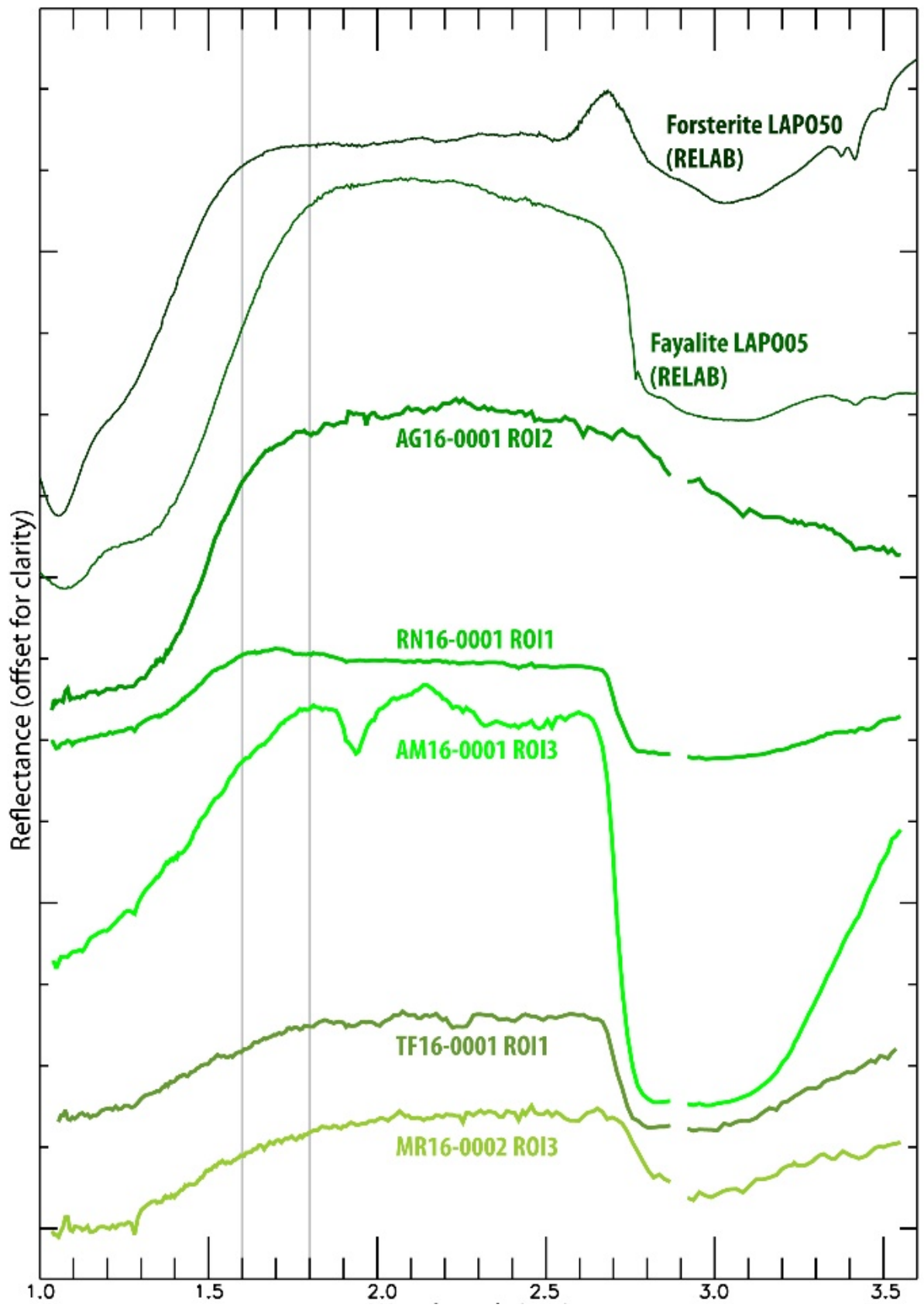

*Figure 10. Selection of local averages of MicrOmega spectra from Canary Islands rock samples compared with olivine reference spectra.*

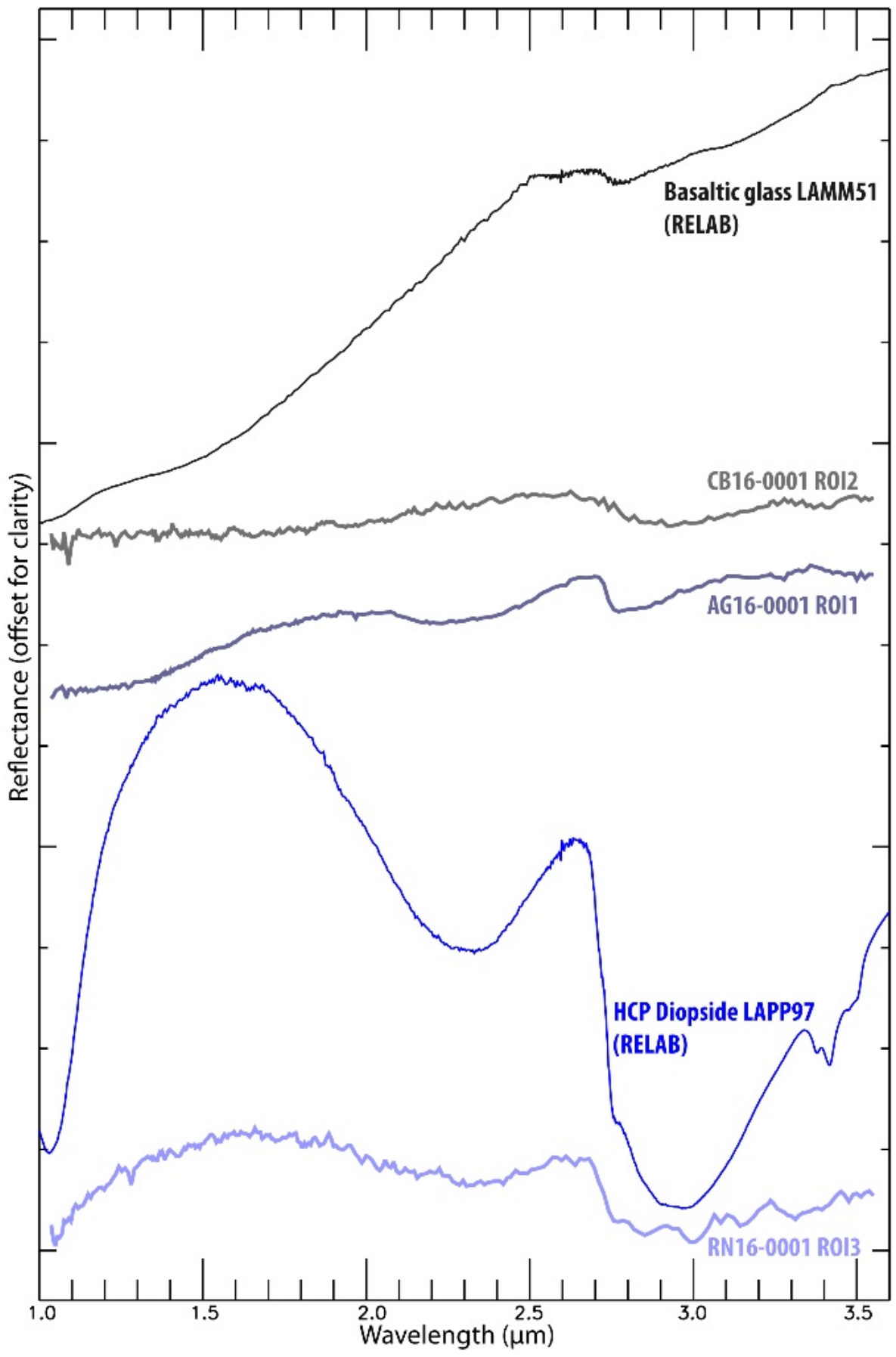

*Figure 11. Selection of local averages of MicrOmega spectra from Canary Islands rock samples compared with reference spectra.*

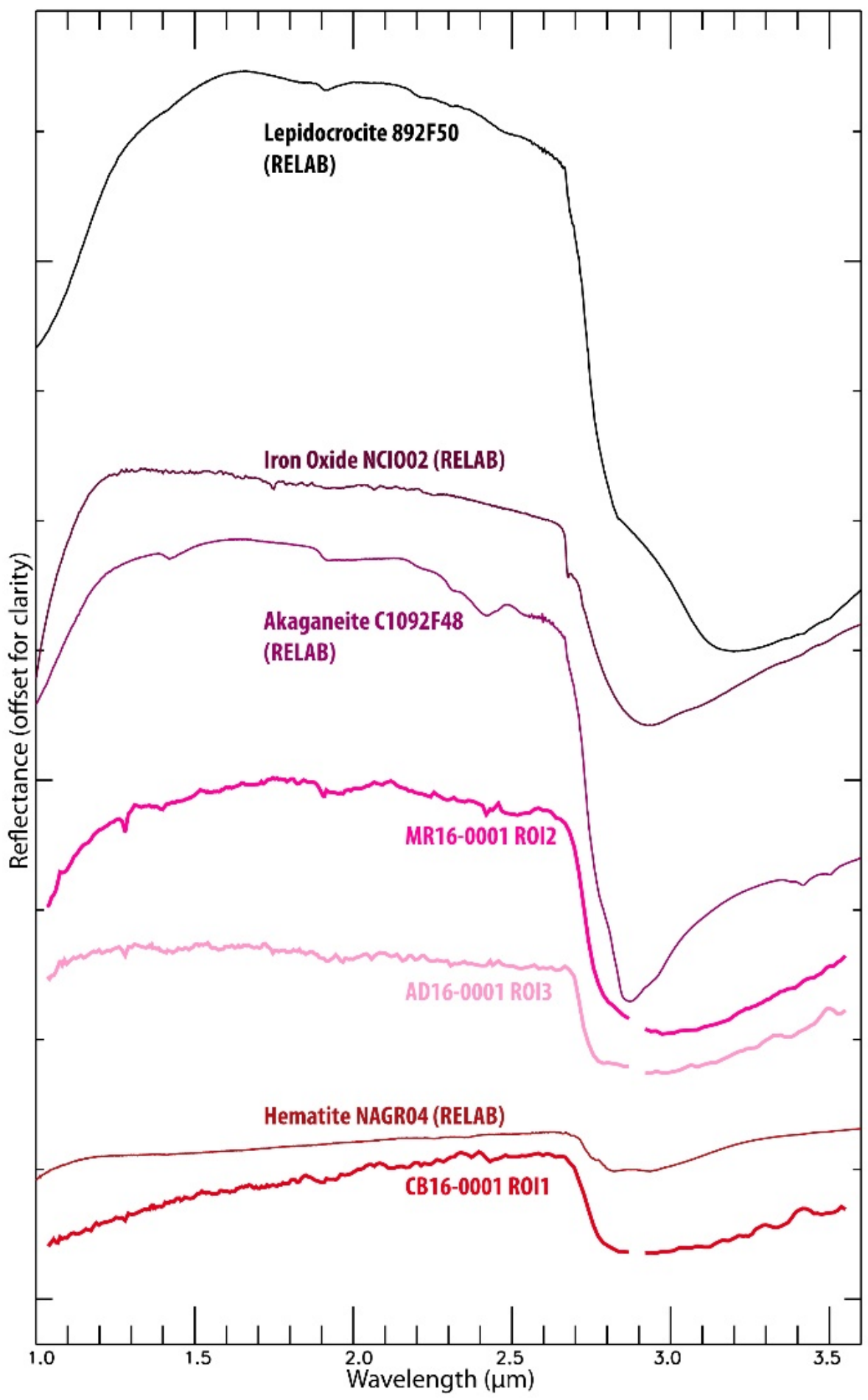

*Figure 12. Selection of local averages of MicrOmega spectra from Canary Islands rock samples compared with reference spectra of oxides.*

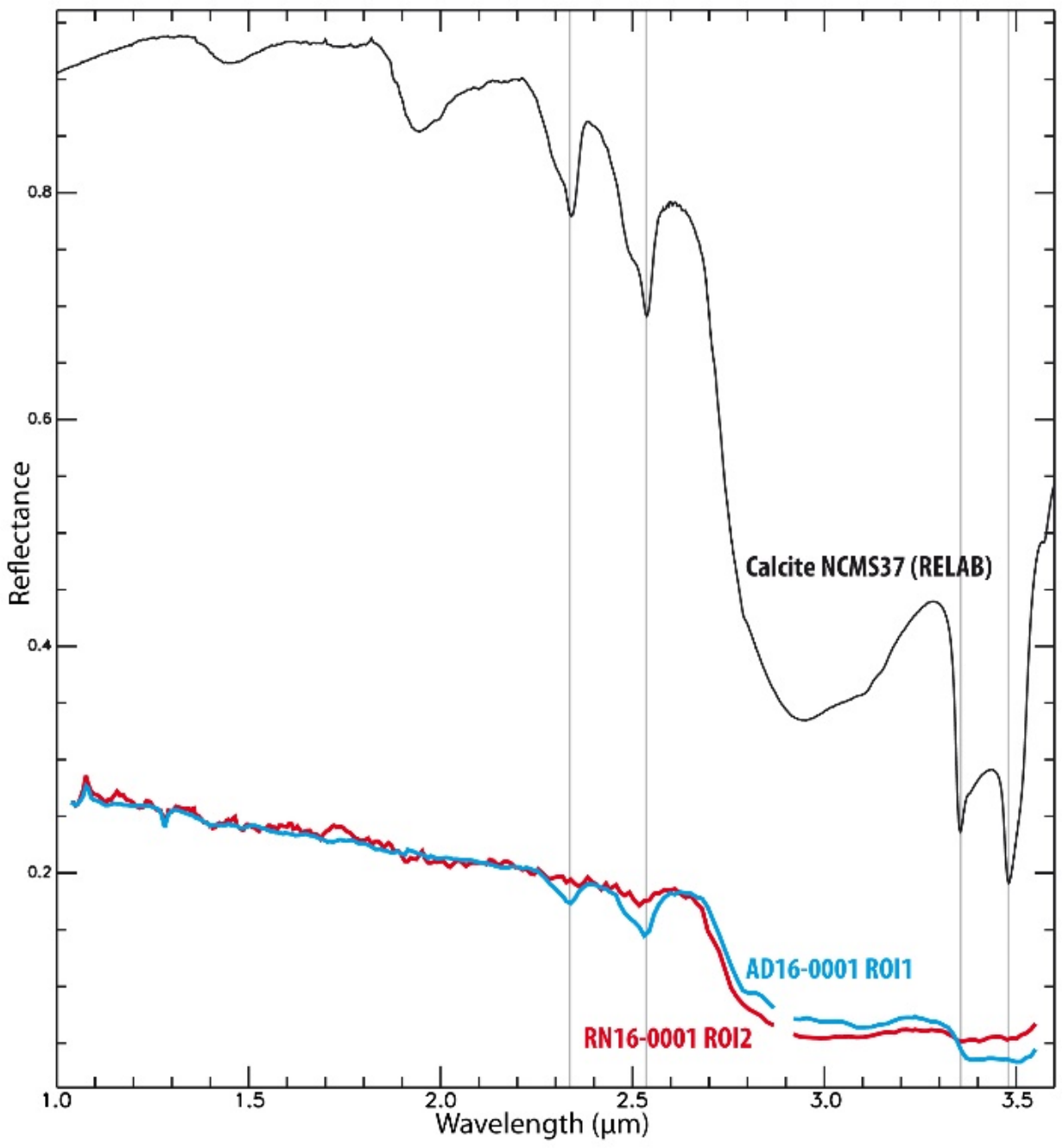

*Figure 13. Selection of local averages of MicrOmega spectra from Canary Islands rock samples compared with a Calcite reference spectrum.*